\documentclass[useAMS,usenatbib]{mn2e}  

\usepackage{amssymb,amsmath}
\usepackage{amsfonts}
\usepackage{graphicx}
\usepackage{booktabs}
\usepackage{tabularx}
\usepackage{color}
\usepackage{ulem}

\numberwithin{equation}{section}

\def\d{{\rm d}}
\def\e{{\rm e}}
\newcommand{\p}{\partial}

\newcommand{\case}{\textstyle\frac}

\newcommand{\vw}{{\mathbf w}}

\newcommand{\vx}{{\mathbf x}}
\newcommand{\vI}{{\mathbf I}}

\newcommand{\suml}{\sum\limits}

\renewcommand{\Re}{\textrm{Re}\,}
\renewcommand{\Im}{\textrm{Im}\,}

\newcommand{\sgn}{{\mathrm{sgn}}}

\begin{document}

\title[Antonov problem reviewed]{Radial orbit instability in systems of highly eccentric orbits: Antonov problem reviewed}
\author[E. V. Polyachenko and I. G. Shukhman]
       {E.~V.~Polyachenko,$^1$\thanks{E-mail: epolyach@inasan.ru},
        I. G. Shukhman,$^2$\thanks{E-mail: shukhman@iszf.irk.ru}\\
       $^1$Institute of Astronomy, Russian Academy of Sciences, 48 Pyatnitskya St., Moscow 119017, Russia\\
       $^2$Institute of Solar-Terrestrial Physics, Russian Academy of Sciences,
       Siberian Branch, P.O. Box 291, Irkutsk 664033, Russia}



\maketitle

\begin{abstract}
Stationary stellar systems with radially elongated orbits are subject to radial orbit instability -- an important phenomenon that structures galaxies. Antonov (1973) presented a formal proof of the instability for spherical systems in the limit of purely radial orbits. However, such spheres have highly inhomogeneous density distributions with singularity $\sim 1/r^2$, resulting in an inconsistency in the proof. The proof can be refined, if one considers an orbital distribution close to purely radial, but not entirely radial, which allows to avoid the central singularity. For this purpose we employ non-singular analogs of generalised polytropes elaborated recently in our work in order to derive and solve new integral equations adopted for calculation of unstable eigenmodes in systems with nearly radial orbits. In addition, we establish a link between our and Antonov's approaches and uncover the meaning of infinite entities in the purely radial case.
Maximum growth rates tend to infinity as the system becomes more and more radially anisotropic. The instability takes place both for even and odd spherical harmonics, with all unstable modes developing rapidly, i.e. having eigenfrequencies comparable to or greater than typical orbital frequencies. This invalidates orbital approximation in the case of systems with all orbits very close to purely radial.
\end{abstract}


\section{Introduction}

The radial orbit instability (ROI), first mentioned in a preprint by \citet{PS72}, plays an important role in the evolution of initially spherically symmetric and axisymmetric systems leading to bar-like perturbations. It has been widely studied both analytically \citep{A73,PS81,PP87,W91,Sa91,P94,PPS11,PPS15} and numerically \citep{P81,M85,M87,B86,AM90,B94,MZ97,TB06}. There are two basic candidates for a physical mechanism of ROI: {\it an analog of Jeans instability} in the anisotropic media and {\it an orbital approach} based on tendency of any pair of orbits to align under their mutual gravity. Discussion on these topics can be found in \citet{PS15}.

A distinct approach is suggested by \citet{MP10}, who give an example of dissipation-induced ROI. A comprehensive modern review on ROI can be found in \citet{MP12}, who also suggest a new symplectic method for exploring stability of equilibrium gravitating systems.

\citet{A73} presents a first formal proof of ROI for purely radial motion using the Lyapunov method. However, his proof is doubtful: the Lyapunov function is ill-defined due to a divergence of its time derivative at the lower limit of integration. Although the main conclusion of the paper is correct, a rigorous examination of the purely radial case is still needed.
The goal of this paper is to reconsider the Antonov problem by applying our technique of an eigenvalue problem in the form of integral equations.

For this purpose, we shall use a general family of models
\begin{equation}
F(E,L)=\,\frac{H(L_T-L)}{L_T^2}\, F_0(E)\ ,
\label{eq:nr}
\end{equation}
where $H(x)$ is the Heaviside function. We retain an arbitrary form for $F_0(E)$ whenever possible, otherwise we admit a polytropic law
\begin{equation}
F_0(E) = \frac{N(L_T,q)}{4\pi^3}(-2E)^q \ .
\label{eq:F0}
\end{equation}
Here $E=\frac{1}{2}(v_r^2+v_{\perp}^2)+\Phi_0(r)\leq 0$ and $L=rv_{\perp}$ are the energy and absolute value of the angular momentum, respectively; $\Phi_0(r)$ is the unperturbed gravitational potential. The additive constant in $\Phi_0$ is chosen so that the potential vanishes at the outer radius of sphere $R$; the normalization constant $N(L_T,q)$ is chosen so that the total mass of the system is $M$. In the calculations below we shall assume that $M=R=G=1$. Equilibrium properties of family (\ref{eq:nr}) with polytropic dependence from energy (\ref{eq:F0}), called {\it softened polytrope models}, are  specially built to consider the limit of purely radial motion and studied in our paper \citep{PPS13}. Stability properties of some series (fixed $q$) are studied in \citet{PS15}.

The polytropic law includes a series of mono-energetic models, in which all stars have zero total energy, at the limit $q\to -1$ \citep[e.g.,][]{GS59}. A limit $L_T \to 0$ in this series gives a well-known \citet{A62} model  which was employed in the Antonov's work. The model is particularly useful  in our case, since it provides the simplest eigenvalue equations, yet preserving all features of interest.

As is already said, our proof is based on analysis and solution of characteristic equations for eigenmodes -- spherical harmonics and corresponding complex frequencies $\omega$, such that ones with the positive imaginary parts give unstable solutions. In Section 2 we derive the equation for a model with {\it purely} radial orbits, using delta-function expansion technique \citep{FP84}. The unperturbed distribution function (DF) of the purely radial system is proportional to the Dirac delta-function of the angular momentum (\ref{eq:pr}), while the perturbed DF is a linear combination of the delta-function and its derivatives (\ref{eq:exp}). The linearised kinetic equation and Poisson equation provide matrix equations (\ref{eq:even_fin}) and (\ref{eq:odd_fin}), for even and odd spherical harmonics, respectively. Both of them contain infinite entities $p_k$ defined by (\ref{eq:pw}) which are a manifestation of the central singularity.

In Section 3 we use the integral equation technique for the two-parametric family of models (\ref{eq:nr}) with {\it nearly} radial orbits \citep{PPS07, PS15}. Since this family includes the purely radial model of Section 2, we can get a link between different parts of the integral equations obtained in Sections 2 and 3, as the control parameter  $L_T$ in the DF approaches zero. In particular, we infer the meaning of the infinite entities, eqs. (\ref{eq:p0lim}, \ref{eq:pk}).

Then, this finding helps us in Section 4 to reduce further the legitimate integral equations of Section 3 for nearly radial orbits to fairly compact limiting integral equations ($L_T \ll 1$), for even and odd spherical harmonics, (\ref{eq:sie1}) and (\ref{eq:sie2}), respectively. They allow to prove existence of the aperiodic even unstable spherical solutions, and absence of the odd unstable spherical solutions. The analytical results are accompanied in Section 5 with numerical eigenmodes' calculation for series $q=-1$ and $q=-1/2$.

In Section 6, we show how the orbital approach breaks down in spherical systems with orbits very close to radial. Comparison of our numerical results with qualitative results by \citet{P91} shows that orbital approach is satisfactory for the systems with orbits of moderate eccentricity only.

Lastly, Section 7 contains a summary and conclusion. Appendix A is devoted to Antonov's `proof' of the existence of ROI (in our terms and notations) with the help of Lyapunov function, as a reminder and demonstration of difficulties appearing in the investigation of the systems with pure radial systems. Appendix B clarifies the sense of diverging coefficients which appear in the equations for purely radial models with the help of limiting procedure from models with finite dispersion over the angular momentum ($L_T\ne 0$).

\section{Pure radial motion: $\delta$-function expansion}

Purely radial orbits possess zero angular momentum, $L=0$. Thus, for systems with purely radial motion, we demand
\begin{equation}
F(E,L)=\delta(L^2)F_0(E)=\frac{\pi}{r^2}\,\delta(v_{\theta})\,\delta(v_{\varphi})\,F_0(E)\ ,
\label{eq:pr}
\end{equation}
where $\delta(x)$ is the Dirac delta-function. The analysis for instability prescribes the following ansatz for the perturbed DF:
\begin{multline}
f_1(t,r,\theta,\varphi,v_r, v_{\theta},v_{\varphi})
 	= A(t,r,\theta,\varphi, v_r)\,\delta(v_{\theta})\,\delta(v_{\varphi})\\
 	+ B(t,r,\theta,\varphi,
v_r)\,\delta'(v_{\theta})\,\delta(v_{\varphi})+
C(t,r,\theta,\varphi,
v_r)\,\delta(v_{\theta})\,\delta'(v_{\varphi})\ , \label{eq:exp}
\end{multline}
where $\delta'$ denotes a derivative of the delta-function. From the linearized kinetic equation
\begin{multline}
\frac{\p f_1}{\p t}+v_r\,\frac{\p f_1}{\p
r}+\frac{v_{\theta}}{r}\,\frac{\p f_1}{\p \theta}+
\frac{v_{\varphi}}{r\,\sin\theta}\, \frac{\p f_1}{\p \varphi}\\+
\left(\frac{v_{\theta}^2+v_{\varphi}^2}{r}-\frac{d\Phi_0}{dr}\right)\,\frac{\p
f_1}{\p v_r}- \left(\frac{v_r\,v_{\theta}}{r}-\cot\theta
\,\frac{v_{\varphi}^2}{r}\right)\,\frac{\p f_1}{\p v_{\theta}} \\
-\left(\frac{v_r\,v_{\varphi}}{r}+\cot\theta\,
\frac{v_{\theta}\,v_{\varphi}}{r}\right)\,\frac{\p f_1}{\p
v_{\varphi}}= \frac{\p \Phi_1}{\p r}\,\frac{\p F}{\p
v_r}\\
+\frac{1}{r}\,\frac{\p \Phi_1}{\p \theta}\,\frac{\p F}{\p fv_{\theta}}+
\frac{1}{r\,\sin\theta}\,\frac{\p\Phi_1}{\p\varphi}\,\frac{\p F}{\p
v_{\varphi}} \ ,
\label{eq:lke}
\end{multline}
relations between decomposition coefficients $A$, $B$, and $C$ can be obtained:
\begin{gather}
\frac{\p A}{\p t}+{\hat D}A+\frac{2 v_r}{r}\,A-\frac{1}{r}\,\frac{\p
B}{\p \theta}-\frac{\cot\theta}{r}\,B
 \nonumber\\-\frac{1}{r\sin\theta}\,\frac{\p C}{\p
\varphi}=\frac{\pi}{r^2}\,F_0'(E)\,v_r\,\frac{\p \Phi_1}{\p r}\ ,\label{eq:s11} \\
\frac{\p B}{\p t}+{\hat D}B+\frac{3
v_r}{r}\,B=\frac{\pi}{r^3}\,F_0(E)\,\frac{\p \Phi_1}{\p \theta}\ , \label{eq:s12} \\
\frac{\p C}{\p t}+{\hat D}C+\frac{3
v_r}{r}\,C=\frac{\pi}{r^3\,\sin\theta}\,F_0(E)\,\frac{\p \Phi_1}{\p
\varphi} \label{eq:s13}
\end{gather}
Here ${\hat D}$ is a differential operator,
\begin{equation}
{\hat D}=v_r\,\frac{\p}{\p r}-\frac{d\Phi_0}{dr}\,\frac{\p}{\p v_r}\ ,
\label{eq:dop}
\end{equation}
and $\Phi_1$ denotes a perturbed potential depending on the polar angle $\theta$ only through Legendre polynomials $P_l$,
\begin{equation}
\Phi_1 \equiv  \chi(t,r)\,P_l(\cos\theta)\ ,
 \label{eq:phi1}
\end{equation}
since the eigenmode spectrum does not depend on the azimuthal number $m$ \citep[e.g.,][]{FP84,B94}.

Substitution to (\ref{eq:s11}--\ref{eq:s13}) gives $C=0$, while $A \propto P_l(\cos\theta)$, and $B \propto \d P_l(\cos\theta)/ \d\theta$. It is convenient to introduce new functions ${\cal A}$ and ${\cal B}$ independent of the angles:
\begin{gather}
A(t,r,\theta,\varphi,v_r)=\frac{{\cal A}(t,r,v_r)}{r^2}\,P_l(\cos\theta)\ \label{eq:AB1a},\\
B(t,r,\theta,\varphi,v_r)=\frac{{\cal B}(t,r,v_r)}{r^3}\,\frac{d
P_l(\cos\theta)}{d\theta}\ .
\label{eq:AB1b}
\end{gather}
The perturbed density
\begin{equation}
\rho_1(t,r,\theta,\varphi) \equiv \Pi(t,r)\,P_l(\cos\theta)
\label{eq:rho1}
\end{equation}
is an integral from ${\cal A}$ over the radial velocity,
\begin{equation}
\Pi(t,r) = \frac{1}{r^2} \int {\cal A}(t,r,v_r)\, dv_r\ .
\label{eq:rhor}
\end{equation}
The eqs. (\ref{eq:s11}--\ref{eq:s13}) and the Poisson equation for the new functions take the form:
\begin{gather}
\frac{\p {\cal A}}{\p t}+{\hat D}{\cal A}+\frac{l\,(l+1)}{r^2}\,{\cal B}=\pi\,v_r\,\frac{d
\chi}{dr}\,F_0'(E)\ ,  \label{eq:AB2a} \\
\frac{\p {\cal B}}{\p t}+{\hat D}{\cal B}=\pi\,\chi(r)\,F_0(E)\ , \label{eq:AB2b} \\
\chi(r)=-\frac{4\pi G}{2l+1}\int dr'\int dv_r'\,{\cal A}(r',v_r')\,{\cal F}_l(r,r')\ ,\label{eq:chia}
\end{gather}
with ${\cal F}_l(r,r')={r_<^l}/{r_>^{l+1}}$, $r_<={\rm min}(r,r')$ and $r_>={\rm max}(r,r')$.

\begin{figure}
  \begin{center}
  \centerline{\includegraphics[width = 85mm]{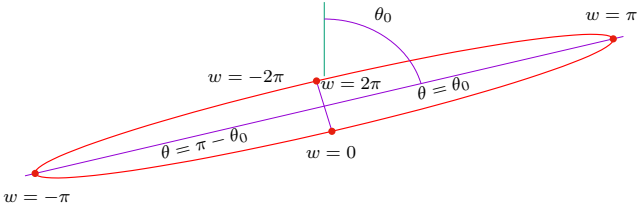}}
  \end{center}
  \caption{A purely radial orbit, $L=0$, as a limiting case of the highly eccentric ellipse. The radial angle variable $w$ is chosen so that $w=0,\pm 2\pi$ correspond to the pericentres, while $w=\pm \pi$ -- to apocentres. Angle $\theta_0$ is the polar angle of the radial orbit, $\theta$ is the polar angle of a star on the radial orbit (the orbit is a spoke in reality; finite thickness here is for illustration of the orbit's variables only).}
  \label{fig1}
\end{figure}

Below, we shall explore the system of eqs. (\ref{eq:AB2a}--\ref{eq:chia}) in terms of action--angle variables. Radial orbits can be treated as highly eccentric ellipses with vanishingly small minor axis, see Fig.\,\ref{fig1}. The stellar position is fixed by four variables, three of which determine the orbit length and orientation (e.g., energy $E$, angles $\theta$ and $\varphi$), and the last one -- radial angle variable $w$ -- sets the position along the orbit,
\begin{equation}
 w=\Omega(E)\int\limits_0^r\frac{dr'}{v_r(E,r')}\ ,
\label{eq:rada}
\end{equation}
where $\Omega(E)$ is the frequency of radial oscillations, $v_r$ is the radial velocity:
\begin{equation}
v_r=\pm \sqrt{2\, [E-\Phi_0(r)]}\ .
\label{eq:vr}
\end{equation}

During full revolution, star's angular variable changes in the range $-2\pi \le w \le 2\pi$. Therefore, the most general functions of $w$ have a period of $4\pi$, and their Fourier expansions should read
\begin{multline}
\{\chi(t,w,E), {\cal A}(t,w,E), {\cal B}(t,w,E) \}  = \\
= \sum\limits_{n=-\infty}^{\infty} \{ \Phi_{n/2}(t,E), A_{n/2}(t,E),
B_{n/2}(t,E)\}\,e^{inw/2}\  \label{eq:fw}
\end{multline}
with $n$ running over all integers. Thus, variables $(t, r, v_r)$ are changed to $(t, E, w)$.

As the star travels from the upper part of the orbit $0<w< 2\pi$ to the lower part $-2\pi < w < 0 $, the polar and azimuthal angles change discontinuously:
\begin{equation}
\theta \to \pi - \theta\ ,\quad \varphi \to \pi + \varphi\ .
\label{eq:angle_jump}
\end{equation}
This results in additional factor in case of odd spherical harmonics, so further analysis should be done separately for even and odd cases.

\subsection{Equations for even spherical harmonics}

In action--angle variables, the evolutionary equations for perturbations in purely radial systems are
\begin{gather}
\frac{\p {\cal A}}{\p t}+\Omega_1\,\frac{\p{\cal A}}{\p w}+\frac{l\,(l+1)}{r^2(E,w)}\,{\cal B}= \pi\,
\Omega_1\,\frac{\p\chi(t, E,w)}{\p w}\,F_0'(E) \ , \label{eq:Aw} \\
\frac{\p{\cal B}}{\p t}+\Omega_1\,\frac{\p{\cal B}}{\p w}=\pi\,\chi(t,E,w)\,F_0(E) \ ,
\label{eq:Bw} \\
\chi(t,E,w)=-\frac{4\pi G}{2l+1}\int \frac{dE'}{\Omega_1(E')}  \nonumber\\
\times \int
dw'\,{\cal A}(t, E',w')\,{\cal F}_l[r(E,w),r'(E',w')]\ .
\label{eq:chiw}
\end{gather}

Radius of a star as a function of angle $w$ obeys the following symmetry conditions:
\begin{equation}
r(2\pi-w) = r(-w) = r(w)\ ,
\label{eq:rw}
\end{equation}
thus coefficients
\begin{multline}
\{ \Phi_{n/2}(t,E), A_{n/2}(t,E), B_{n/2}(t,E)\}  \\
= \frac{1}{4\pi}
\int\limits_{-2\pi}^{2\pi}  \{\chi(t,r), {\cal A}(t,r,v_r), {\cal
B}(t,r,v_r) \} \,e^{-inw/2}\,dw \label{eq:chin}
\end{multline}
vanish for odd $n$, and equal to
\begin{multline}
 \{ \Phi_{k}(t,E), A_{k}(t,E), B_{k}(t,E)\}   \\
= \frac{1}{\pi} \int\limits_{0}^{\pi}  \{\chi(t,r), {\cal A}(r,v_r),
{\cal B}(r,v_r) \} \,\cos(kw)\,dw \, ,
\label{eq:chin_even}
\end{multline}
otherwise ($k=n/2$). In this case, periodicity changes to $2\pi$ due to symmetry of potential and density perturbations with respect to transformation $w \to -w$.

Assuming that perturbations are $\propto \exp(-i\omega t)$, one can obtain from (\ref{eq:Aw})--(\ref{eq:chiw})
\begin{gather}
-i\,(\omega-k\,\Omega)\,A_k+l\,(l+1)\!\!\!\sum\limits_{k'=-\infty}^{\infty}
\!\!p_{k-k'}B_{k'}=i\pi\,k\,\Phi_k\,F_0'(E)\ ,\label{eq:Aw1} \\
-i\,(\omega-k\,\Omega)\,B_k=\pi\,\Phi_k\,F_0(E)\quad \label{eq:Bw1}
\end{gather}
and
\begin{equation}
\Phi_k(E)=-\frac{2G}{2l+1}\int\frac{dE'}{\Omega(E')}
\sum\limits_{k'=-\infty}^{\infty}{\cal K}_{k\,k'}^{\rm even}
(E,E')\,A_{k'}(E')\ , \label{eq:Pw1}
\end{equation}
where $k$ and $k'$ are integers,
\begin{gather}
p_k(E)=\frac{1}{2\pi}\oint \frac{dw}{r^2(E,w)}\,e^{-ikw}=\frac{1}{2\pi}\oint \frac{\cos (kw)}{r^2(E,w)}\,dw
\label{eq:pw}
\end{gather}
and
\begin{equation}
{\cal K}_{k\,k'}^{\rm even}(E,E')=4
\int\limits_{0}^{\pi}dw\int\limits_{0}^{\pi}dw'\,
\cos(kw)\,\cos(k'w')\,{\cal F}_l(r,r')\ .
\end{equation}

Eq. (\ref{eq:pw}) emphasises an issue arising in systems with purely radial orbits -- the integrals diverge in the centre ($w \to 0, |2\pi|$). Thus, these expressions for $p_k$ require an interpretation. Note that a similar difficulty appeared in \citet{A73}, but then no adequate attention has been paid. For example, $\varpi(E)\equiv p_0(E)$ is $1/r^2$, averaged along the orbit:
\begin{equation}
p_0=\Big\langle \frac{1}{r^2}\Big\rangle\equiv\frac{1}{2\pi}\oint \frac
  {dw}{r^2}=\frac{\Omega}{\pi}\int\limits_0^{r_{\rm
  max}(E)}\frac{dr}{r^2\,\sqrt{2E-2\Phi_0(r)}}\ ,
\label{eq:p0}
\end{equation}
and diverges evidently at $r=0$, since singularity of $\Phi_0$ is weaker than $1/r^2$. We plan to tackle the issue employing a family of models with nearly radial orbits and study the system of interest by considering more and more radially anisotropic systems (see Section 3).

With (\ref{eq:Aw1}) and (\ref{eq:Bw1}), one can exclude $A_k$ and $B_k$ from the equation in favour of $\Phi_k$:
\begin{multline}
\Phi_k(E)=-\frac{2\,\pi\,G}{2l+1}\int\frac{dE'}{\Omega(E')}
\sum\limits_{k'}\,\frac{{\cal
K}_{k\,k'}^{\rm even}(E,E')}{\omega-n'\,\Omega(E')}\  \\
\times\left[l\,(l+1)\,
 F_0(E')\,\sum\limits_m \frac{p_{k'-m}(E')\,\Phi_m(E')}{\omega-m\,\Omega(E')}  \right. \\
\left. - \Omega(E')\,k'\,\Phi_{k'}(E')\,\dfrac{dF_0(E')}{dE'}\right]\ .
\label{eq:Phin_eu}
\end{multline}

For some $F_0(E)$ (e.g., polytropes (\ref{eq:F0}) with $q<0$) the integral from the term including $dF_0/dE'$ diverges. In this case one should use the Lagrangian form, which is obtained formally by integration by parts and omission of the surface term:
\begin{multline}
\Phi_n(E)=-\frac{2\,\pi\,G}{2l+1}\int dE'\,F_0(E') \\
\times \sum\limits_{n'}\left\{\frac{l\,(l+1)}{\Omega(E')}\,\frac{ {\cal K}_{n\,n'}^{\rm
even}(E,E')}{\omega-n'\,\Omega(E')}
 \sum\limits_m \frac{p_{n'-m}(E')\,\Phi_m(E')}{\omega-m\,\Omega(E')} \right. \\
 \left. + \frac{d}{dE'}\left[
\,n'\,\Phi_{n'}(E')\,\frac {{\cal K}_{n\,n'}^{\rm
even}(E,E')}{\omega-n'\,\Omega(E')}\right]\right\}\
\label{eq:Phin_lg}
\end{multline}
(see \citet{PS15} for details). For further analysis, it is convenient to use an alternative form of the last equation. Using the identity provided $m\ne n'$
\begin{equation}
\dfrac{1}{\omega\!-\!n'\,\Omega}\cdot\dfrac{1}{\omega\!-\!m\,\Omega}
\!=\!\frac{1}{\Omega\,(n'\!-\!m)}
\left(\frac{1}{\omega\!-\!n'\,\Omega}\!-\!\frac{1}{\omega\!-\!m\,\Omega}\right),
\label{eq:id1}
\end{equation}
one can have
\begin{multline}
\Phi_n(E)=-\frac{2\,\pi G\,l\,(l+1)}{2l+1}\int
\frac{dE'\,F_0(E')}{\Omega(E')}\\
\times \sum\limits_{n'}{\cal K}_{n\,n'}^{\rm even}(E,E')\,\Bigg\{
 \sum\limits_{m\ne n'}
 \frac{p_{n'-m}(E')\,\Phi_m(E')}{\Omega(E')\,(n'-m)}\,
  \\
\times\left[\frac{1}{\omega-n'\,\Omega(E')} - \frac{1}{\omega-m\,\Omega(E')} \right]
 +\frac{\Phi_n'(E')\,p_0(E')}{[\omega-n'\,\Omega(E')]^2}\Bigg\} \\
-\frac{2\,\pi G}{2l+1}\int dE'\,F_0(E')\sum\limits_{n'}
\frac{d}{dE'}\left[ \,n'\,\Phi_{n'}(E')\,\frac {{\cal
K}_{n\,n'}^{\rm even}(E,E')}{\omega-n'\,\Omega(E')}\right]\ .
\label{eq:ierad}
\end{multline}
In the final equation, we have separated the last term which retains even in the case of radial oscillations $l=0$.

To find out the meaning of integrals (\ref{eq:pw}) in expressions for $p_k$, a specific model is not important, since $p_k(E)$ depends on the orbit, but not on the orbit distribution over the phase space. For simplicity we consider a monoenergetic model corresponding to $q\to -1$ limit in (\ref{eq:F0}), which leads to algebraic equations:
\begin{multline}
\Phi_k=Q_L^{\rm pure\ radial}+Q_E^{\rm pure\ radial}\\ \equiv -\frac{1}{4\pi^2}\,\frac{l\,(l+1)}{2l+1}
\sum\limits_{k'}{\cal K}_{k\,k'}^{\rm even}(0,0)\,\Bigg[
 \sum\limits_{m\ne k'}
 \frac{p_{k'-m}\,\Phi_m}{\Omega\,(k'-m)}\,  \\
 \times  \left(\frac{1}{\omega-k'\,\Omega} - \frac{1}{\omega-m\,\Omega} \right)
 +\frac{\Phi_{k'}\,p_0}{(\omega-k'\,\Omega)^2}\Bigg]
 \\
-\frac{1}{4\pi^2}\,\frac{\Omega}{2l+1}\left[\frac{d}{dE'}\sum\limits_{k'}
\,k'\,\Phi_{k'}(E')\,\frac {{\cal K}_{k\,k'}^{\rm
even}(0,E')}{\omega-k'\,\Omega(E')}\right]_{E'=0}\ ,
\label{eq:even_fin}
\end{multline}
where $\Phi_k$ now denotes $\Phi_k(E=0)$.

\subsection{Equations for odd spherical harmonics}

The jump of the polar angle (\ref{eq:angle_jump}) gives rise to additional factor
\begin{equation}
\sigma(w) = \sgn\bigl[\sin({\case{1}{2}}\,w)\bigr]\ ,
\label{eq:add_f}
\end{equation}
in case of the odd spherical harmonics, i.e.
\begin{gather}
\Phi_1 =\chi(t,r)\,\sigma(w)\,P_l(\cos\theta)\ , \\
\rho_1(t,r,\theta,\varphi) =\Pi(t,r)\,\sigma(w)\,P_l(\cos\theta)\ .
\label{eq:phi_rho1_odd}
\end{gather}
The functions to be expanded
\begin{gather}
\bar \chi(t,w) \equiv \chi(t, r)\,\sigma(w)\ , \\
\bar {\cal A}(t,w,E) \equiv {\cal A}(t,w,E)\,\sigma(w)\
\label{eq:phi_rho1_odd_exp}
\end{gather}
are antisymmetric, i.e.
\begin{equation}
\bar \chi(t,-w) = -\bar \chi(t,w)\ ,\quad \bar {\cal A}(t,-w,E) = -\bar {\cal A}(t,w,E)\ .
\label{eq:as}
\end{equation}
Now the expansion coefficients for even $n$ vanish, while for odd $n$ one has ($n=2k+1$):
\begin{multline}
\{ \Phi_{k+1/2}(t,E),  A_{k+1/2}(t,E)\}  \\
=\frac{1}{i\pi}
\int\limits_{0}^{\pi}  \{\chi(t,r),  {\cal A}(t,r,v_r)\}
\,\sin[(k+{\case{1}{2}})\,w]\,dw \ .
\label{eq:chin_odd}
\end{multline}

From the eqs. similar to (\ref{eq:Aw}) and (\ref{eq:Bw}), it follows that expansion coefficients $B_{n/2}(t,E)$ also vanish for even $n$. For a new set of variables $\{\bar\Phi_{k}, \bar A_{k},\bar B_{k}\} \equiv \{\Phi_{k+1/2}, A_{k+1/2},B_{k+1/2}\}$, one obtains equations for the odd spherical harmonics $l$:
\begin{gather}
-i\,[\omega-(k+{\case{1}{2}})\,\Omega]\, \bar
A_k+l\,(l+1)\,\sum\limits_{k'=-\infty}^{\infty}p_{k-k'}
\bar B_{k'} \nonumber \\
=i\pi\,(k+1/2)\,\bar \Phi_k\,F_0'(E)\ ,\label{eq:Aw1Odd} \\
-i\,[\omega-(k+{\case{1}{2}})\,\Omega]\,\bar B_k=\pi\,\bar \Phi_k\,F_0(E)\quad \label{eq:Bw1Odd}
\end{gather}
and
\begin{equation}
\bar \Phi_k(E)=-\frac{2G}{2l+1}\int\frac{dE'}{\Omega(E')}
\sum\limits_{k'=-\infty}^{\infty} {\cal K}_{k\,k'}^{\rm
odd}(E,E')\,\bar A_{k'}(E')\ , \label{eq:odd_puasson}
\end{equation}
where $k$ and $k'$ are integers; $p_k(E)$ is given by (\ref{eq:pw}), i.e. the same as for the even $l$;
\begin{multline}
{\cal K}_{k\,k'}^{\rm odd}(E,E')\\=4
\int\limits_{0}^{\pi}dw\int\limits_{0}^{\pi}dw'\,
\sin[(k+{\case{1}{2}})\,w]\,\sin[(k'+{\case{1}{2}})\,w']\,{\cal
F}_l(r,r')\ .
\end{multline}
Eliminating $\bar A_{k}(E)$ and $\bar B_{k}(E)$ in favour of $\bar \Phi_{k}(E)$, one obtains the equations similar to
(\ref{eq:even_fin}) for even $l$:
\begin{multline}
\bar\Phi_k=-\frac{1}{4\pi^2}\,\frac{l\,(l+1)}{2l+1}
\sum\limits_{n'}{\cal K}_{k\,k'}^{\rm odd}(0,0) \Bigg\{
 \sum\limits_{m\ne k'}
 \frac{p_{k'-m}\,\bar\Phi_m}{\Omega\,(k'-m)}\, \\
\times
 \left[\frac{1}{\omega-(k'+{\case{1}{2}})\,\Omega}
 -\frac{1}{\omega-(m+{\case{1}{2}})\,\Omega} \right]
 +\frac{\bar\Phi_{k'}\,p_0}{[\omega-(k'+{\case{1}{2}})\,\Omega)^2}\Bigg\}
 \\
-\frac{1}{4\pi^2}\,\frac{\Omega}{2l+1}\left[\frac{d}{dE'}\sum\limits_{k'}
\,(k'+{\case{1}{2}})\,\Phi_{k'}(E')\,\frac {{\cal K}_{k\,k'}^{\rm
odd}(0,E')}{\omega-(k'+{\case{1}{2}})\,\Omega(E')}\right]_{E'=0}\ .
\label{eq:odd_fin}
\end{multline}

\section{Nearly radial orbits: integral equations}

The integral equations (\ref{eq:even_fin}) and (\ref{eq:odd_fin}) are of no use, since they contain infinite coefficients $p_k$.
In this section we consider nearly radial series of `dispersed' Agekyan models ($q=-1$) with a control parameter $L_T$ which includes the purely radial (Agekyan) model of the previous section as a limiting case $L_T \to 0$. We shall see, that the dispersed models allow for a well-defined integral equations, and their solutions indeed give infinitely large growth rates in the limit of the purely radial case.

The integral equations for the nearly radial models in the Lagrangian form are \citep{PS15}:
\begin{multline}
    \phi_{\,l_1,\,l_2}(E,L)\!=\!-\frac{4\pi
    G}{2l+1}\!\sum\limits_{l_1'=-\infty}^{\infty}\sum\limits_{l_2'=-l}^l\!\!
    D_l^{l_2'}\!\! \int\!\! dE'\!\!\int\!\! dL'\,F(E',L') \\
 \times \left[\dfrac{\p }{\p E'}\,\Omega_{l_1'l_2'}(E',L')
        + l_2'\,\dfrac{\p
        }{\p L'}\,\right] \frac{L'}{\Omega_1(E'\!\!,L')}\\
    \times\frac{\phi_{\,l_1'\,l_2'}(E',L')
\Pi_{l_1,\,l_2;\,l_1',\,l_2'}(E,L;E',L')}
    {\omega-\Omega_{l_1'l_2'}(E',L')}\ ,
\label{eq:ie}
\end{multline}
where $\Omega_{l_1'l_2'}(E',L')\equiv l_1'\,\Omega_1(E',L') + l_2'\,\Omega_2(E',L')$; differentiation operator $\p/\p E'$ acts both on $\Omega_{l_1'l_2'}(E',L')$ and the last row; coefficients $D_l^k$ vanish for odd $|l-k|$ and
$$
 D_l^k=
  \dfrac{1}{2^{2\,l}}
  \,\dfrac{(l+k)!(l-k)!}{\Bigl[\bigl(\frac{1}{2}\,(l-k)\bigr)!\,
 \bigl(\frac{1}{2}\,(l+k)\bigr)!{\phantom{\big|}}\Bigr]^2}
$$
otherwise. For the given models, the right-hand side can be written as the sum of two terms:
\begin{multline}
\phi_{l_1l_2}(0,L_T) = Q_L+Q_E\\
\equiv-\frac{{\bar K(L_T)}}{2\pi^2(2l+1)\,L_T}\sum\limits_{l_1'=-\infty}^{\infty}
\sum\limits_{l_2'=-l}^l (l_2'D_l^{l_2'})\\
\times \frac{\phi_{l_1'l_2'}(0,L_T)\,\Pi_{l_1,l_2;l_1'l_2'}(0,L_T;0,L_T)}{\Omega_1(0,L_T)\,
[\omega-\Omega_{l_1'l_2'}(0,L_T)]}\\
-\frac{K(L_T)}{2\pi^2\,(2l+1)\,L_T^2}\!\!\sum\limits_{l_1'=-\infty}^{\infty}
\sum\limits_{l_2'=-l}^l\!\! D_l^{l_2'}
\Biggl[\frac{\p}{\p E'}\int\limits_0^{L_T}\!\!
L'dL'
\frac{\Omega_{l_1'l_2'}(E',L')}{\Omega_1(E',L')}\\
\times \frac{\phi_{l_1'l_2'}(E',L')\,\Pi_{l_1,l_2;l_1'l_2'}(0,L;E',L')}
{\omega-\Omega_{l_1'l_2'}(E',L')}\,\Biggr]_{E'=0} \ .
\label{eq:ieagk}
\end{multline}
Due to orbit symmetry, $r(w) = r(-w)$, the kernel functions $\Pi_{l_1,\,l_2;\,l_1',\,l_2'}$ and unknown expansion coefficients of the potential $\phi_{l_1\,l_2}$ can be expressed in integral forms with integration reduced from $[-\pi,\pi]$ to $[0,\pi]$:
\begin{multline}
    \Pi_{l_1,\,l_2;\,l_1',\,l_2'}(E,L;E',L')\\
 =\oint dw \cos\Theta_{l_1\,l_2}(w)\oint dw'\cos\Theta_{l_1'\,l_2'}(w') \,{\cal
    F}_l(r,r')\\ =4\int\limits_{0}^{\pi} dw \cos\Theta_{l_1\,l_2}(w)\int\limits_0^{\pi} dw'\cos\Theta_{l_1'\,l_2'}(w')\,
     \,{\cal F}_l(r,r')\ ;
\label{eq:Pi}
\end{multline}
\begin{equation}
\phi_{l_1\,l_2}(E,L)=\frac{1}{\pi}\int\limits_0^{\pi}\cos\Theta_{l_1
l_2}(E,L;w)\, \chi\bigl[r(E,L,w)\bigr]\,dw\ .
\label{eq:phi}
\end{equation}
The angle  $\Theta_{l_1l_2}(E,L,w)$ is
\begin{equation}
\Theta_{l_1\,l_2}(E,L;w)= \bigl(l_1+l_2\,\frac{\Omega_2}{\Omega_1}\bigr)\,w -l_2\delta\varphi(E,L;w)\ ,
\label{eq:Theta}
\end{equation}
with
\begin{multline}
\delta\varphi(E,L,w) =L\int\limits_{r_{\rm min}(E,\,L)}^{r(E,L,w)}
\frac{dx}{x^2\,\sqrt{\phantom{\big|}[2E+2\Psi(x)]-L^2/x^2}}\\
=\frac{L}{\Omega_1}\int\limits_0^w
\frac{dw'}{r^2(w')}
\label{eq:dphi}
\end{multline}
denoting the azimuthal change of the particle coordinate as it passes from the pericentre to the current radius $r$; $\Psi(r) \equiv - \Phi_0(r)$. At the apocentre, $\delta\varphi(E,L;\pi)=(\Omega_2/\Omega_1)\,\pi $.

Further, we shall expand the functions entering eq. (\ref{eq:ie}) considering $L_T$ as a small parameter. For nearly radial orbits, the precession velocity
\begin{equation}
\Omega_{\rm pr} \equiv \Omega_2-\frac12 \,\Omega_1
\label{eq:Om_pr}
\end{equation}
is small compared to frequencies $\Omega_{1,2}$. So, it can be separated out in the linear combination
\begin{equation}
    \Omega_{l_1l_2}=l_1\Omega_1+l_2\Omega_2=
    (l_1+{\case{1}{2}}\,l_2)\,\Omega_1+l_2\,\Omega_{\rm pr}\ .
\end{equation}
The angle $\Theta_{l_1\,l_2}(E,L;w)$ can be written as a sum
\begin{equation}
    \Theta_{l_1\,l_2}(E,L;w)=
    [(l_1+{\case{1}{2}}\,l_2)\,w-{\case{1}{2}}\,l_2\pi]+l_2\,\beta\ ,
\label{eq:L46}
\end{equation}
where the expression in the square brackets retains in the limit $L\to 0$, while
\begin{multline}
    \beta=\frac{\Omega_{\rm pr}}{\Omega_1}\,(w-\pi)+\frac{L}{\Omega_1}\int\limits_w^{\pi}\frac{dw'}{r^2(w')} =\frac{\Omega_{\rm pr}}{\Omega_1}\,(w-\pi)  \\
	  +L\int\limits_r^{r_{\rm max}} \frac{dr'}{r'^2\,\sqrt{\phantom{\big|}[2E+2\Psi(r')]-L^2/r'^2}}
\label{eq:L47}
\end{multline}
vanishes.

In Appendix B, we give details of the expansion of eq. (\ref{eq:ie}) in $L_T$ and $\beta$ for the even spherical harmonics $l$. It should be compared with eq. (\ref{eq:even_fin}) for the systems with purely radial orbits. The equations coincide entirely, if $p_0(E)$ is substituted by the limiting ratio $\Omega_{\rm pr}(L_T)/L_T$, i.e.
\begin{equation}
    p_0 \equiv \frac{1}{2\pi}\oint \frac{dw}{r^2(w)}
    \to \lim\limits_{L_T\to  0} \frac{\Omega_{\rm pr}(L_T)}{L_T}\ ,
\label{eq:p0lim}
\end{equation}
and coefficients $p_k(E)$ for $k\ne 0$ are understood as the limits
\begin{multline}
 p_k(E)\equiv
\frac{1}{2\pi}\oint \frac{\cos(kw)}{r^2(w)}\,dw  \\
\to \lim\limits_{L_T\to 0}
\frac{1}{L_T}\Bigl[\frac{L_T}{\pi}\int\limits_0^{\pi}
\frac{\cos(kw)}{r^2}\,dw-\frac{\Omega_1(L_T)}{2}\Bigr]\, .
\label{eq:pk}
\end{multline}
Given that $(L_T/\pi)\int_0^{\pi} dw/{r^2}=\Omega_2$, one can have
\begin{equation}
    p_k-p_0 =
    \frac{2}{\pi}\int\limits_0^{\pi} \frac {\sin^2 (\frac{1}{2}\,k
    w)}{r^2(w)}\,dw\ ,
\label{eq:pklim}
\end{equation}
where the right-hand side converges in the usual sense. Note that expansion of (\ref{eq:ie}), not given here, and comparison with (\ref{eq:odd_fin}) for the odd harmonics lead to the same results (\ref{eq:p0lim}) and (\ref{eq:pk}).

The obtained relation between $p_k$ and the limiting value of $\varpi \equiv \Omega_{\textrm {pr}}(L_T)/L_T$ implies that $p_k$ are infinitely large. Indeed, in purely radial systems the density is {\it necessarily singular}, at least not weaker than $1/r^2$ \citep{BJ68, RT84}. Thus the potential and gravitational force are also singular and linear law of the precession rate $\Omega_{\textrm pr}(L)$ is no longer valid. In particular, for the softened polytropes all purely radial models ($q \leq 1/2$) give $\varpi \to \infty$ in the limit $L_T \to 0$ \citep[see Fig.9b in][]{PPS13}. Besides, for dispersed Agekyan model, we found numerically that $\Omega_{\rm pr}(L_T)\approx 0.316\,(L_T)^{0.26}$, i.e., $\varpi \approx 0.316/(L_T)^{0.74}$ (see below Sect. 5.1).

\section{Limiting integral equations}

Eqs. (\ref{eq:p0lim}) and (\ref{eq:pklim}) for $p_k(E)$ show that infinitely large coefficients occur in (\ref{eq:Aw1})--(\ref{eq:Pw1}) and (\ref{eq:Aw1Odd})--(\ref{eq:odd_puasson}) as $L_T$ goes to zero. This enables us to obtain a simplified counterparts of the stability equations. We shall start from the equations in the form (\ref{eq:Aw})--(\ref{eq:chiw}), and assume everywhere that $L_T\ll 1$. The right hand side in (\ref{eq:Aw}) should be omitted since it does not contain $p_0$. Then, one should neglect the difference between $p_k$ and $p_0=\varpi \equiv \Omega_{\textrm {pr}}(L_T)/L_T \gg 1$ since $p_k/p_0 -1 = {\cal O}(1/p_0) \ll 1$. The expansion $1/r^2=\sum p_k\,\exp (ikw)$ then turns into
\begin{equation}
\frac{1}{r^2} \approx 2\pi\,p_0\,\sum_n \delta(\,w-2\,\pi\,n)\ ,
\label{eq:1r2}
\end{equation}
so that (\ref{eq:Aw}) and (\ref{eq:Bw}) turn into
\begin{gather}
\frac{\p {\cal A}}{\p t}+\Omega\,\frac{\p{\cal A}}{\p w}=-2\pi\,\varpi\,l\,(l+1)\,\delta(w)\,{\cal B}\equiv {\cal R}\ ,
\\
\frac{\p{\cal B}}{\p t}+\Omega\,\frac{\p{\cal B}}{\p w}=\pi\,\chi(w)\,F_0(E)\ .
\end{gather}
Now changing $\p/\p t$ to $-i\omega$ and solving the equations, taking into account symmetry of functions ${\cal R}(w)$ and $\chi(w)$, one obtains for ${\cal A}$:
\begin{multline}
{\cal A}(E,w) = \frac{e^{\,i\,\nu \,w}}{\Omega(E)}\,\left[\frac{1}{1-e^{\,2i\pi\,\nu}}
{\displaystyle\int_{-\pi}^{\pi}dw\, {\cal R}(w)\,e^{-i\,\nu\,w}}\right. \\
-\left.\int_w^{\pi} dw'{\cal R}(w')\,e^{-i\,\nu\,w'}\right]\ ,
\end{multline}
or
\begin{equation}
{\cal A}=-\frac{i\pi\,
l\,(l+1)\,\varpi}{\Omega\,\sin(\pi\,\nu)}\,{\cal
B}(0) \exp [i \nu (w- \pi\,\sgn w)]  \ ,
\end{equation}
where $\nu \equiv \omega/\Omega$ is the dimensionless frequency. For ${\cal B}$ the solution is
\begin{multline}
{\cal B}(E,w)=\frac{\pi F_0(E)}{\Omega(E)}\,e^{\,i\,\nu \,w}\left[\frac{1}{1-e^{2i\pi\,\nu}}
\int_{-\pi}^{\pi}dw\, \chi(w)\,e^{-i\,\nu\,w}\right.\\
\left.-\int_w^{\pi} dw'\chi(w')\,e^{-i\,\nu\,w'}\right].
\end{multline}
In particular, ${\cal B}(E, 0)$ is
\begin{equation}
{\cal B}(E, 0)=\frac{i\,\pi\,F_0(E)}{\Omega\,\sin\bigl(\pi\,\nu\bigr)} \Phi_\omega(E)\ ,
\end{equation}
where
\begin{equation}
\Phi_\omega(E)\equiv  \int\limits_0^{\pi} \chi \,\cos\bigl[\nu\,(w-\pi)\bigr]\,dw\ .
\end{equation}
Multiplying the Poisson equation (\ref{eq:chiw}) by  $\cos\bigl[\nu\,(w-\pi)\bigr]$
and integrating from $0$ to $\pi$, one obtains an integral equation for even $l$:
\begin{multline}
\Phi_{\omega}(E)=-\frac{8\pi^3
G\,l\,(l+1)}{2l+1}\int\frac{dE'}{(\Omega')^3\,\sin^2\bigl(\pi\,\nu'\bigr)}\,
F_0(E')\\
\times \varpi(E',L_T)\,{\cal
K}_{\omega}(E,E')\,\Phi_{\omega}(E')\ ,
\label{eq:even}
\end{multline}
where $\Omega'$ and $\nu'$ denote $\Omega(E')$ and $\omega/\Omega'$, and the kernel is
\begin{equation}
{\cal K}_{\omega}(E,E')\!=\!\!\int\limits_0^{\pi}
\!\!\cos\bigl[\nu\,(w-\pi)\bigr]\,dw\!\!\int\limits_0^{\pi}\!\!
\cos\bigl[\nu'\,(w'-\pi)\bigr]\,dw'\,{\cal
F}_l(r,r')\ .
\label{eq:kern}
\end{equation}

The analogous equation of odd $l$ has the form:
\begin{multline}
\Phi_{\omega}(E)=-\frac{8\pi^3
G\,l\,(l+1)}{2l+1}\int\frac{dE'}{(\Omega')^3\,\cos^2\bigl(\pi\,\nu'\bigr)}\,
F_0(E') \\
\times\varpi(E',L_T)\,{\cal
K}_{\omega}(E,E')\,\Phi_{\omega}(E')\ .
\label{eq:odd}
\end{multline}

We shall refer further to eqs. (\ref{eq:even}) and (\ref{eq:odd}) as {\it limiting integral equations}. Note that they lack the advantage of the linear eigenvalue problem, since frequency $\omega$ enters into the kernel function and into the argument of sine and cosine in denominators. However, they retain their forms during the change $\omega \to -\omega$, so both equations should depend on $\omega^2$.

We need to emphasise that $\varpi$ depends on $L_T$, and that it is assumed that $L_T \ll 1$ and $\varpi(E,L_T) \gg 1$. This is the only variable dependent on $L_T$, in all other places $L_T \to 0$ limit leads to finite quantities, so there we assume $L_T=0$.

Relative simplicity of the limiting integral equations (\ref{eq:even}) and (\ref{eq:odd}) allows us to demonstrate analytically  existence of {\it aperiodic} unstable solutions ($\omega = i\gamma$ with $\gamma>0$) for even $l$ and their absence for odd $l$. Introducing $\sigma(E) \equiv \gamma/\Omega(E)$ one obtains equivalent equations for $\gamma$:

\medskip\noindent for even $l$,
\begin{multline}
\Phi_{\gamma}(E)=\frac{8\pi^3
G\,l\,(l+1)}{2l+1}\int\frac{F_0(E')\,\varpi(E',L_T)\,dE'}{(\Omega')^3\,
\sinh^2(\pi\,\sigma')}\, \\
\times{\cal K}_{\gamma}(E,E')\,\Phi_{\gamma}(E')\quad \textrm{and}
\label{eq:sie1}
\end{multline}

\medskip\noindent for odd $l$,
\begin{multline}
\Phi_{\gamma}(E)=-\frac{8\pi^3
G\,l\,(l+1)}{2l+1}\int\frac{F_0(E')\,\varpi(E',L_T)\,dE'}{(\Omega')^3\,
\cosh^2(\pi\,\sigma')}\,\\
\times{\cal K}_{\gamma}(E,E')\,\Phi_{\gamma}(E')\ .
\label{eq:sie2}
\end{multline}
In both equations
\begin{multline}
{\cal K}_{\gamma}(E,E')=\int\limits_0^{\pi}
\cosh\bigl[\sigma\,(w-\pi)\bigr]\,dw \\
\times\int\limits_0^{\pi}
\cosh\bigl[(\sigma'\,(w'-\pi)\bigr]\,dw'\,{\cal F}_l(r,r')\ .
\label{eq:sie_kg}
\end{multline}
Redefinition of the eigenfunction
$$\Psi_{\gamma}(E) = \frac{\Phi_{\gamma}(E)}{\sinh(\pi\sigma)}\,\sqrt
{\frac{F_0(E)\,\varpi(E)}{\Omega^3(E)}}$$ allows one to symmetrize the integral equations. Using an integral representation for ${\cal F}_l(r,r')$ through Bessel functions \citep[e.g.,][eq. 6.574]{GR15},
\begin{equation}
 {\cal F}_l(r,r')=(2l+1)\int_0^{\infty} dk\,
 \frac{J_{l+1/2}(kr)}{\sqrt{kr}}\,\frac{J_{l+1/2}(kr')}{\sqrt{kr'}}\ ,
\label{eq:fie}
\end{equation}
it can be proven that the kernel of the symmetrized equation for even $l$, ${\cal Q}_{\gamma}^{\rm even}$ is {\it positive}. So the eigenvalue problem (\ref{eq:sie1}) can be rewritten as
\begin{equation}
\int dE'\,{\cal Q}_{\gamma}^{\rm
even}(E,E')\,\Psi_{\gamma}(E')=\Lambda_n(\gamma)\,\Psi_{\gamma}(E)\ .
\label{eq:integr_even}
\end{equation}
Here $\Lambda_n(\gamma)$ $(n=0,1,2,...)$,  are a set of positive eigenvalues of the linear problem depending on $\gamma$ as a parameter. The eigenvalues $\Lambda_n$ can be ordered so that $\Lambda_0>\Lambda_1>\Lambda_2>...$, and larger $n$ correspond to eigenfunctions with larger number of nodes ($n=0$ eigenfunction has the largest scale). The needed values of $\gamma$ satisfy
\begin{equation}\Lambda_n(\gamma)=1\ .
\label{eq:lam}
\end{equation}

In the limit $\gamma\to 0$ frequencies $\sigma$ and $\sigma'$ are vanishingly small, and the kernel
\begin{multline}
{\cal Q}_{\gamma}^{\rm even}(E,E')\stackrel{\gamma\to 0}{\approx}\frac{8\pi
G\,l\,(l+1)}{2l+1}\,\sqrt{\frac{F_0(E)\,\varpi(E)}{\Omega^3(E)}}\, \\ \times
\sqrt{\frac{F_0(E')\,\varpi(E')}{\Omega^3(E')}}\frac{1}{\sigma\,\sigma'}\int\limits_0^{\pi}
dw\int\limits_0^{\pi} dw'\,{\cal F}_l(r,r')\gg 1
\label{eq:g-small}
\end{multline}
is large, so many $\Lambda_n$ are greater than 1. On the other hand, for $\gamma\gg 1$
\begin{equation}
\sinh(\sigma\,\pi)\approx {\case{1}{2}}\,e^{\sigma\,\pi}\ ,\ \
\cosh[\sigma\,(w-\pi)]\approx {\case{1}{2}}\,e^{\sigma\,(\pi-w)}\ ,
\label{eq:scmu}
\end{equation}
and the kernel takes a form:
\begin{multline}
{\cal Q}_{\gamma}^{\rm even}(E,E')\stackrel{\gamma\to \infty}{\approx}\frac{8\pi^3
G\,l\,(l+1)}{2l+1}\,\sqrt{\frac{F_0(E)\,\varpi(E)}{\Omega^3(E)}}\,  \\
\times
\sqrt{\frac{F_0(E')\,\varpi(E')}{\Omega^3(E')}} \int\limits_0^{\pi} e^{-\sigma\,w}\,dw\int\limits_0^{\pi}
e^{-\sigma'\,w'}\,dw'\,{\cal F}_l(r,r')\ .
\label{eq:ie_large_g}
\end{multline}
Due to rapidly decreasing exponents the kernel is small, and thus $\Lambda_n$ are small. When $\gamma$ is changing from zero to infinity, many $\Lambda_n$ cross the unity value. Since for a given $\gamma$, the eigenfunction with the largest scale has the largest eigenvalue, $\Lambda_0$ will be the first to cross unity as $\gamma$ increases, so $\gamma_0 > \gamma_1 > \gamma_2 > ...\,$.

An equation analogous to (\ref{eq:integr_even}) for odd $l$ has a {negative} kernel, ${\cal Q}_{\gamma}^{\rm odd} < 0$. It means that all $\Lambda_n$  are negative for any $\gamma$, and no aperiodic solution is possible.

\section{Numerical results}

\subsection{Aperiodic modes in the dispersed Agekyan model ($q=-1$)}

For the dispersed Agekyan model
\begin{equation}
F_0(E)=\dfrac{\Omega}{8\pi^3}\,\delta(E)
\end{equation}
which corresponds to $q=-1$ in (\ref{eq:F0}) the integral equation (\ref{eq:sie1}) is reduced to an algebraic one,
\begin{equation}
\frac{l\,(l+1)}{2l+1}\,\frac{\varpi(0,L_T)}{\Omega^2}\,\frac{{\cal
K}_{\gamma}(0,0)}{\sinh^2(\pi\,\sigma)}=1\ ,
\label{even_q1}
\end{equation}
where $\sigma = {\gamma}/{\Omega}$,
\begin{multline}
{\cal K}_{\gamma}(0,0)=\int\limits_0^{\pi}
\cosh\bigl[\sigma\,(w-\pi)\bigr]\,dw \\
\times\int\limits_0^{\pi}
\cosh\bigl[(\sigma\,(w'-\pi)\bigr]\,dw'\,{\cal F}_l(r,r')\ .
\end{multline}
It is known \citep{PS15} that this equation has only one even aperiodic solution $\gamma$, which is large when $\varpi$ is large. Thus, keeping in the hyperbolic functions the leading exponents only, one obtains the characteristic equation for even $l$ aperiodic modes,
\begin{multline}
\frac{l\,(l+1)}{2l+1}\,\frac{\varpi(0,L_T)}{\Omega^2}  \\ \times \,\int\limits_0^{\pi}\
dw\int\limits_0^{\pi} dw' \,\exp
\bigl[-\frac{\gamma}{\Omega}\,(w+w')\bigr]\,{\cal F}_l(r,r')=1\ .
\label{eq:q1_even}
\end{multline}

\begin{figure}
   \centerline{\includegraphics[width = 85mm]{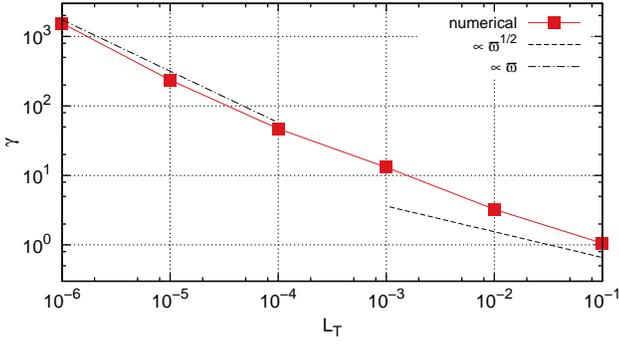}}
  \caption{Growth rates of the aperiodic eigenmodes in the Agekyan model (spherical harmonics $l=2$).}
  \label{fig3}
\end{figure}

For this model, the function $\varpi(L_T)$ can be approximated by the power law
\begin{equation}
  \varpi(L_T)\simeq\frac{0.316}{L_T^{\ 0.74}}\ ,
  \label{eq:L308}
\end{equation}
obtained numerically in the range $-3<\lg L_T<-1$. With this approximation formula, one can find solutions at arbitrary small $L_T$. Fig.\,\ref{fig3} shows the dependence of the growth rate $\gamma$ of the unstable aperiodic solution for $l=2$. As expected, $\gamma$ is large as $L_T \to 0$  and it scales approximately as $\gamma \sim \varpi$ for very small $L_T$, and $\gamma \sim \varpi^{1/2}$ for $L_T \sim 0.1$.  Recall that the dynamic frequency $\Omega=2.16$ is of the order unity, thus the obtained growth rates obey the inequality $|\gamma| > \Omega$ for $L_T < 10^{-2}$.

\subsection{Oscillatory modes in the Agekyan model}

This approximation formula (\ref{eq:L308}) allows us to calculate oscillatory unstable solutions in the form of even and odd spherical harmonics using
\begin{equation}
-\frac{l\,(l+1)}{2l+1}\,\frac{\varpi(L_T)}{\Omega^2}\,\frac{{\cal
K}_{\omega}(0,0)}{\sin^2(\pi\,\nu)}=1
  \label{eq:L309}
\end{equation}
for even $l$, and
\begin{equation}
-\frac{l\,(l+1)}{2l+1}\,\frac{\varpi(L_T)}{\Omega^2}\,\frac{{\cal
K}_{\omega}(0,0)}{\cos^2(\pi\,\nu)}=1
  \label{eq:L310}
\end{equation}
for odd $l$, where $\nu = {\omega}/{\Omega}$ and
\begin{multline}
{\cal K}_{\omega}(0,0)=\int\limits_0^{\pi} dw\,
\cos\bigl[\nu\,(w-\pi)\bigr]  \\ \times \int\limits_0^{\pi} dw'\,
\cos\bigl[(\nu\,(w'-\pi)\bigr]\,{\cal F}_l(r,r')\ .
  \label{eq:L311}
\end{multline}

\begin{figure}
  \centerline{\includegraphics[width = 85mm]{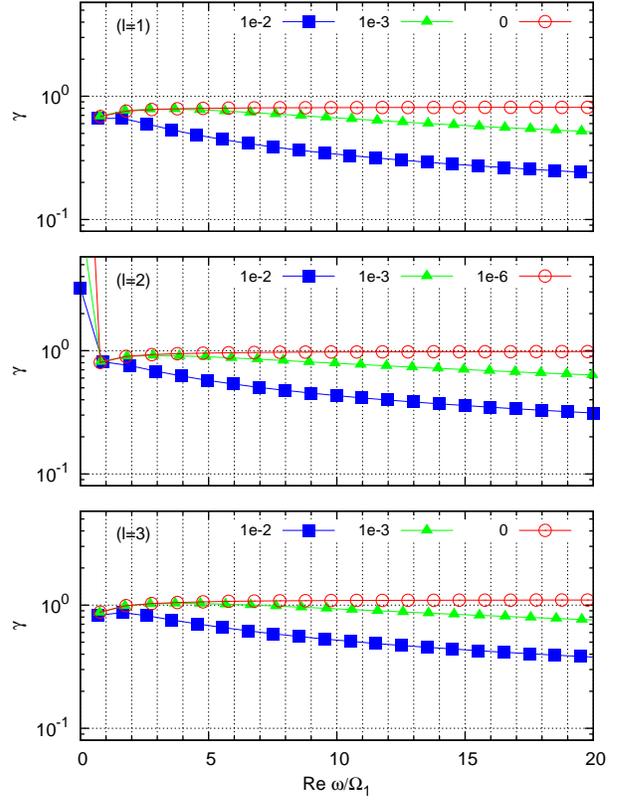}}
  \caption{Oscillatory unstable modes $\Re\omega \ne 0$ for the dispersed Agekyan model: (a) $l=1$ spherical harmonics, (b) $l=2$ and (c) $l=3$.}
  \label{fig4}
\end{figure}

The results for the first three harmonics in a wide range of frequencies and $\lg L_T = -2 ... -6$ are presented in panels of Fig.\,\ref{fig4}. As expected, the aperiodic solutions are absent for odd modes. The growth rates of aperiodic solutions (for $l=2$) rapidly increase as $L_T \to 0$, so for most values of $L_T$ the apriodic solutions are outside the (middle) panel.

Growth rates of the oscillatory solutions show weak dependence on $\Re\omega$, especially for the smallest $L_T$ when $\gamma = \Im\omega \approx 1$. Real parts of frequencies obey approximately $\Re\omega /\Omega = n-1/4$ in the limit $L_T \to 0$, but these limiting values approach from different sides (in case of even $l$ -- from the right, and in case of odd $l$ -- from the left). In all cases the oscillatory solutions obey $|\omega| \gtrsim \Omega$.

\subsection{Series $q=-1/2$}

In this section we study a series of models with nontrivial dependence of the DF on the energy. For $q=-1/2$ the potential can be obtained in an analytical form for arbitrary $L_T$ \citep{PPS13} and this explains our choice of $q$.

\begin{figure}
  \centerline{\includegraphics[width = 85mm]{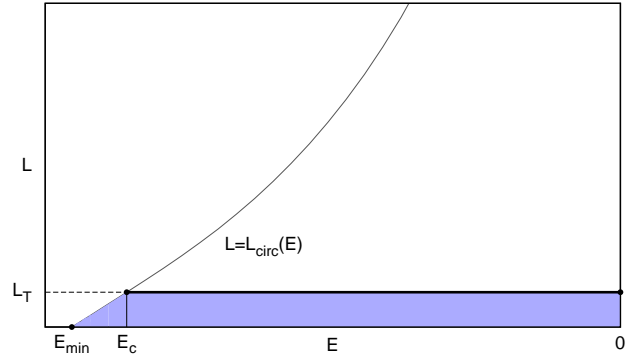}}
  \caption{The phase space in the $q=-1/2$ series (filled area), and the line of integration $E_c<E<0$, $L=L_T$ (thick line) in the integral equations (\ref{eq:even}) and (\ref{eq:odd}).}
  \label{fig:phase}
\end{figure}

In the purely radial limit $\Phi = \ln r$, $\Omega = \sqrt{2\pi} e^{-E}$. If $L_T$ is small but finite, there is a small radius $r_1 = {\cal O}(L_T)$ which separates two intervals. From $r_1$ to approximately 1,
the potential is close to $\Phi = \ln r$, but in the interval $[0, r_1]$ it behaves like $-\sin{kr}/r$, with $k \propto L_T^{-1}$. However, for energies in the range $[E_c, 0]$, where $E_c=E_c(L_T)$ is the energy of the particle on the circular orbit with angular momentum $L_T$ (see Fig.\,\ref{fig:phase}), the pericentre distance is not less than $r_1$, i.e. one can use the potential for the purely radial case to calculate the precession rate.
The azimuth change $g(\alpha)$ during the pericentre passage is
\begin{equation}
  g (\alpha) = \pi +\case {1} {2} \, \pi\mu \, \bigl (1 +\case {1} {2} \, \mu \,
  \ln 2\mu\bigr) + {\cal O} \left (\mu^3\ln^2\mu\right)\ ,
\end{equation}
where $\alpha \equiv L/L_{\rm circ}(E)$, $L_{\rm circ}(E) = e^{E-1/2}$, $\mu = [\ln\left(1/\alpha\right)]^{-1}$
  \citep{TT97, PPS13} and thus the precession rate of the particle with energy $E$ and angular momenta $L$ is
\begin{equation}
  \Omega_{\textrm {pr}}(E,L)\approx \frac{e^{-E}}{\sqrt{8\pi}}\,\mu\,\bigl (1
  +\case{1}{2}\,\mu\,\ln 2\mu\bigr)\ ,
  \label{eq:ompr_q12a}
\end{equation}
where
$$
\mu=\bigl[\,\ln\left(1/\alpha\right)\bigr]^{-1}=\bigl[\ln (e^{E-1/2}/L)\bigr]^{-1}\gg 1\ .
$$
However, this formula is valid for nearly radial orbits $E>E_c$ only and fails for the circular orbit $E=E_c$ where $\Omega_{\textrm {pr}}=0.293/L_T$. So, instead of  (\ref{eq:ompr_q12a}), we shall calculate the precession rate numerically, using $\Phi = \ln r$ for the potential, and
\begin{equation}
  \Omega_{\textrm {pr}}(E,L_T)=\,\frac{g(\alpha_T)-\pi}{2h(\alpha_T)}\,e^{-E} ,
  \label{eq:ompr_q12b}
\end{equation}
where
\begin{equation}
  g(\alpha)=\frac{2\alpha}{\sqrt{e}}
  \int\limits_{x_{\rm min}}^{x_{\rm max}}\dfrac{dx}{x\,\sqrt{-2x^2\,\ln x-\alpha^2/e}}
  \label{eq:ompr_q12_g}
\end{equation}
and
\begin{equation}
  h(\alpha)=\int\limits_{x_{\rm min}}^{x_{\rm max}}\dfrac{x\,dx}{\sqrt{-2x^2\,\ln x-\alpha^2/e}}\ .
  \label{eq:ompr_q12_h}
\end{equation}
With new variables $z=e^{E}$, $z_c=\sqrt{e}\,L_T$ and $\nu=\omega/\Omega_1(z,L_T)$, where
 \begin{equation}
  \Omega_1(z,L_T)=\frac{\pi}{z\,h(z_c/z)}\ ,
  \label{eq:om_q12}
\end{equation}
the limiting integral equations (\ref{eq:even}) and (\ref{eq:odd}) become
\begin{multline}
  \Phi(z)=-\sqrt{\frac{1}{2\pi^3}}\,
  \frac{l(l+1)}{(2l+1)}\int_{z_c}^1\frac{z'^2\,\varpi(z',L_T)\,dz'}
  {\sqrt{\ln\bigl(1/z'^2\bigr)}} \, \\ \times\,
  \frac{{\cal K}_{\omega}(z,z')}{\sin^2(\pi\nu')}\,\Phi(z')
  \label{eq:even_q12}
\end{multline}
for even $l$, and
\begin{multline}
  \Phi(z)=-\sqrt{\frac{1}{2\pi^3}}\,\frac{l(l+1)}{(2l+1)}
  \int_{z_c}^1\frac{z'^2\,\varpi(z',L_T)\,dz'}{\sqrt{\ln\bigl(1/z'^2\bigr)}} \\ \times\,
  \,\frac{{\cal K}_{\omega}(z,z')}{\cos^2(\pi\nu')}\,\Phi(z')\ ,
  \label{eq:odd_q12}
\end{multline}
for odd $l$, where
\begin{equation}
  \varpi(z,L_T)=\frac{\Omega_{\rm pr}(z,L_T)}{L_T}\ ,
  \label{eq:p0_q12}
\end{equation}
and the kernel is given by eq. (\ref{eq:kern}).

\begin{figure}
  \centerline{\includegraphics[width = 85mm]{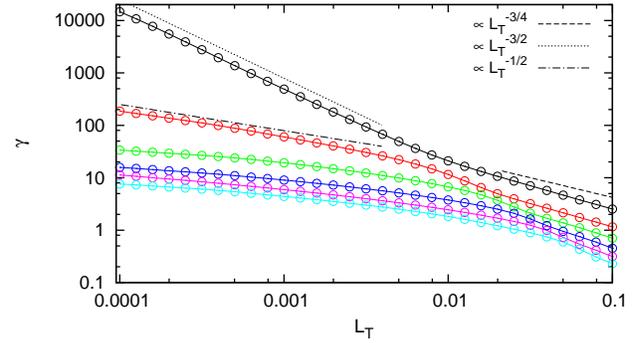}}
  \caption{Growth rates of the  first six aperiodic eigenmodes in the $q=-1/2$ series as functions of $L_T$ (spherical harmonics $l=2$).}
  \label{fig-aper12}
\end{figure}

The instability growth rates of the aperiodic modes $\omega = i\gamma$ ($l=2$) are presented in Fig.\,\ref{fig-aper12}. Contrary to the Agekyan model (Fig.\,\ref{fig3}), in this case we have many aperiodic solutions.

\begin{figure}
  \centerline{\includegraphics[width = 85mm]{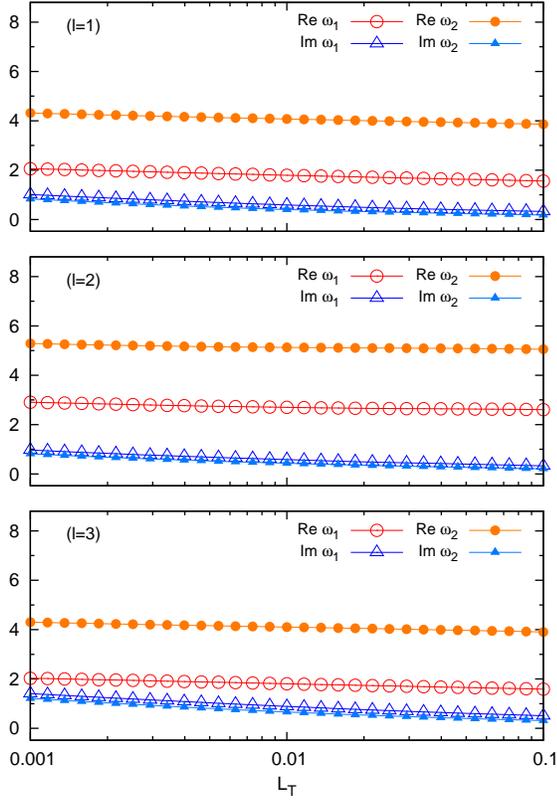}}
  \caption{Two oscillatory unstable solutions as functions of $L_T$ for $l=1, 2, 3$.}
  \label{fig5}
\end{figure}

Results of our calculations of oscillatory eigenmodes for $l=1...3$ spherical harmonics are presented in Fig.\,\ref{fig5}. Panels of the figure show both real and imaginary parts of $\omega$ of the first two modes versus the control parameter $L_T$. The difference between two successive real parts is $\approx 2.3$, and the overall behaviour resembles one of the oscillatory modes in the dispersed Agekyan model (see Fig.\,\ref{fig4}).

\section{The orbital approach in systems with nearly radial stellar orbits}

In this section we shall analyse validity of the orbital approach in studying systems with purely radial and nearly radial orbits. Recall that the orbital approach turns from consideration of a particle trajectory to precessing motion of the closed orbital wires. An angle between two successive apocentres of the particle on the radial orbit in the scale free potentials $\Phi \propto r^{s}$ is
\begin{equation}
\delta\varphi  = \left\{
\begin{array}{rl}
\pi \ , & s \geq 0\ , \\
\displaystyle \frac{2\pi}{2 + s} \ , & s < 0 \ 	
\end{array} \right.
\end{equation}
\citep{TT97}. The potentials of the softened polytropes in the limit of purely radial motion $L_T \to 0$ diverge in the centre as \citep{PPS13}
\begin{equation}
\Phi \propto \ln^p (1/r)\quad \textrm{and} \quad p = (1/2 - q)^{-1}\ ,
\end{equation}
i.e. weaker than any negative power $s$, so $\delta\varphi = \pi$, and according to (\ref{eq:dphi}), $\Omega_1(E,L=0) = 2\Omega_2(E,L=0)$. The precession rate of the nearly radial orbits $L \ll 1$,
\begin{equation}
\Omega_{\textrm pr} = \Omega_2 - \frac 12 \Omega_1
\end{equation}
is slow, $\Omega_{\textrm pr} \ll \Omega_{1,2}$.

Now consider the motion of a particle on a nearly radial orbit in presence of a weak non-rotating slowly growing bar potential
\begin{equation}
H = H_0 + \epsilon \Phi_{\textrm b}(\vx,t) \ .
\label{eq:iham}
\end{equation}
Switching to the action--angle variables in the orbital plane, $\vx = \vx(\vI, \vw)$, the small perturbation due to the bar can be written as the Fourier series over radial angle $w_1$ of the unperturbed orbit:
\begin{equation}
\Phi_{\textrm b}(\vx, t) = \e^{\gamma t} \suml_{l} \Phi_l(\vI) \e^{i l w_1 + i m w_2}\ ,\ \  m=2.
\end{equation}
The phases $l w_1 + m w_2 = (l\Omega_1 + m\Omega_2)t$ vary quickly for all $l$ except $l=-1$. Thus, omitting quickly oscillating terms one obtains an `averaged' hamiltonian
\begin{equation}
\overline H = H_0 + \epsilon \Phi_{-1}(\vI) \e^{i m \overline w_2 + \gamma t}\ ,
\label{eq:aham}
\end{equation}
which possesses an adiabatic invariant $J \equiv I_2 + I_1/2$ and `slow' angle variable $\overline w_2 =w_2-\frac{1}{2}\,w_1$ \citep{LB79, EP04}. The equations of motion are
\begin{align}
\dot J & = -\frac{\p \overline H}{\p w_1} = 0\ , \label{eq:a_em1} \\
\dot I_2 & = -\frac{\p \overline H}{\p \overline w_2} = -i m \epsilon \Phi_{-1}(\vI) \e^{i m \overline w_2 + \gamma t}\ , \label{eq:a_em2} \\
\dot{w}_1 & = \Omega_1 + \epsilon \left. \frac{\p \Phi_{-1}}{\p J} \right|_{I_2}  \e^{i m \overline w_2 + \gamma t}\quad \textrm{and} \label{eq:a_em3} \\
\dot{\overline w}_2 & = \Omega_{\textrm pr} + \epsilon \left. \frac{\p \Phi_{-1}}{\p I_2}  \right|_{J}  \e^{i m \overline w_2 + \gamma t}\ .\label{eq:a_em4}
\end{align}
In particular, eq. (\ref{eq:a_em2}) gives the change of the angular momentum perpendicular to the orbital plane, $I_2$, and eq. (\ref{eq:a_em4}) describes the apsidal precession. The requirement of adiabaticity implies
\begin{equation}
\gamma \ll \Omega_1 \ .
\label{eq:ad}
\end{equation}

\citet{P91} obtained an expression for the growth rate in the monoenergetic model ($q=-1$) with pure radial orbits in the framework of spoke approximation, when the orbital wires turn into spokes. In our notations the growth rate is
\begin{equation}
\gamma^2\equiv-\omega^2=\frac{l(l+1)\,\Omega^2\varpi}{\pi^2}\int\limits_0^{\infty} dk\,I_l^2(k)\ ,
\label{gr_spoke}
\end{equation}
where $\Omega$ is the radial frequency,
$$
I_l(k)=\int\limits_0^1{dr}\,\rho_{\rm lin}(r)\,\frac{J_{l+1/2}(kr)}{\sqrt{kr}}\ ,
$$
$\rho_{\rm lin}(r) =1/|v_r|= 1/\sqrt{2\Psi(r)}$ is a linear density of the spoke; $\varpi \equiv \bigl[d\Omega_{\rm pr}/dL\bigr]_{L=0}$. However, as we argued in Sections 3 and 5, the last parameter grows infinitely, as we turn to more and more radially anisotropic systems.

It is interesting to note that if we {\it formally} assume the scaled growth rate of the mode $\sigma=\gamma/\Omega$ to be small in Eq. (\ref{even_q1}) for monoenergetic model ($q=-1$), and use identity (\ref{eq:fie}) for ${\cal F}_l(r,r')$, we obtain exactly the same expression for the growth rate (\ref{gr_spoke}) found by \citet{P91} in the spoke approximation. This fact justifies the spoke approximation for systems with sufficiently small $\varpi$ (moderately elongated orbits), but not for the very eccentric orbits! Note also that the growth rate $\gamma$ for $q=-1$ series scales as $\varpi^{1/2}$ for not too small $L_T$ in agreement with (\ref{gr_spoke}), but for very small $L_T$ grows even faster than $\varpi$.

Hence stability study of the spherical systems with nearly radial or purely radial orbits cannot be made in the framework of the orbital approach (and the spoke approximation in particular), since $\gamma$ grows with $\varpi$ and condition (\ref{eq:ad}) fails.

\section{Summary and Conclusions}

Using a new technique based on integral eigenvalue equations, we reconsider here a well-known work on radial orbit instability by Antonov (1973) in which spherical models with purely radial motion are studied. The Antonov problem cannot be correctly solved in the purely radial models due to singularity in the centre. Thus series of models with parameter $L_T$ controlling orbit eccentricity including purely radial model (corresponding to $L_T=0$) should be used.

The derived integral equations involve an only large quantity $\varpi \equiv \Omega_{\textrm{pr}}(L_T)/L_T$ in case of small $L_T$ tending to infinity as $L_T$ goes to zero. This quantity coincides with the Lynden-Bell derivative $[\p \Omega_{\textrm{pr}}/ \p L]_{L=0}$ playing a crucial role in theory of radial orbit instability \citep{LB79}.

We investigated stability of the spherically symmetric models with respect to perturbations $\propto \chi(r)\,P_l(\cos\theta)$ and obtained numerical solutions for two series of softened polytropic models $F(E,q) \propto H(L_T-L) (-2\,E)^q$ ($H(x)$ is the Heaviside function) allowing the purely radial limit \citep{PPS13, PS15}.

The first one, $q=-1$, is a dispersed Agekyan model. The instability exists both for even and odd spherical harmonics $l$, for which multiple oscillatory modes with $\Re\omega \approx (n-1/4)\,\Omega$ are found, where $n=1,2, ...$; $\Omega \approx 2.16$ is the radial frequency of particles (in units $G=M=R=1$). The modes growth rates $\gamma \equiv \Im\omega \approx 1$. Besides, we found aperiodic modes $\Re\omega=0$ for even spherical harmonics with growth rates tending to infinity as $L_T \to 0$.

The second series $q=-1/2$ provides an analytic potential for any value of parameter $L_T$, and relatively simple formulae for the radial frequency and the precession rate for nearly radial models. As with the previous series, we found multiple oscillatory modes for even and odd spherical harmonics. A characteristic feature of this model is multiple aperiodic modes with growth rates increasing as $L_T \to 0$. We conclude that in all cases (both series, aperiodic and oscillatory modes, even and odd $l$) $|\omega|$ values are of the order of or larger than $\Omega$.

There are several interpretations for the physical mechanism of radial orbit instability (ROI). One relates ROI to the well-known Jeans instability in anisotropic medium for which insufficient velocity dispersion perpendicular to the radial direction cannot resist gravitational clusterization \citep{PS72}. Another one is connected to precession dynamics of eccentric orbits that attract to each other, provided $[\p \Omega_{\textrm{pr}}/ \p L]_{L=0}>0$ \citep{LB79, M87}. This point of view can be justified only for the so-called `slow modes' which satisfy `slow' integral equation in which only one resonance term $\propto [2\Omega_{\textrm{pr}} - \omega]^{-1}$ is retained \citep{P94}. In turn, this implies (i) even $l$ only, and `slowness' of the mode, i.e. $|\omega|$ should be much less than the dynamical frequencies, e.g. $\Omega$ \citep{PS15}. As we saw, none of the solutions obtained in this work satisfy any of these requirements, and we must conclude that the orbital interpretation is limited.

Using the energy approach, \citet{MP10} argue that instability in sufficiently anisotropic systems can be induced by dissipation inevitably present in the real stellar systems. The energy approach claims that if the second order variation of energy due to the perturbation, $H^{(2)}$,  is negative, then the system may be unstable. If, in addition, a small dissipation takes place, the system is guaranteed to be unstable, with the growth rate proportional to the dissipation. Note, however, that the energy approach makes no conclusions for systems without dissipation in the case of negative sign of $H^{(2)}$. In other words, it is of little help for highly anisotropic spherical systems subject to very strong collisionless (i.e., non-dissipative) radial orbit instability, which is apparently more important than the instability potentially induced by dissipation.

Similar to \citet{A73}, we consider here non-radial perturbations independent of parity $l$, and an instability mechanism independent of the suggestion of slowness. However, strong singularity of central density inherent to the system with purely radial orbits \citep{BJ68, RT84} leads to singularity of the potential, consequently infinite $\varpi$ and the growth rates $\gamma$ for even aperiodic modes. We suppose that this instability is manifestation of Jeans instability modified due to periodic radial motion of stars along their orbit.

It is worth recalling, in this context, the argument against our interpretation of ROI, raised for the first time by \citet{M87}. According  to the virial theorem, the growth rate of Jeans instability is of the same order as the inverse crossing time, $\sim(G\rho)^{1/2}$. Since radially anisotropic systems are also strongly radially inhomogeneous, he claims that ``unstable mode would scarcely begin to grow before the particles contributed to it had moved away from their initial positions, to regions of very different density and velocity dispersion''. The  growth rates obtained in our calculations, however, are large compared to the inverse crossing time, which prevent particles from being escaped before the instability takes over the system. Thus, the virial estimate and the entire argument are  not valid for the systems with orbits very close to purely radial.

\section*{Acknowledgments}

We thank Dr. J. Perez for his comments when reviewing the paper, and Dr. R. Moetazedian for his help in improving the English language. This work was supported by the Sonderforschungsbereich SFB 881 ``The Milky Way System'' (subproject A6)
of the German Research Foundation (DFG), and by the Volkswagen Foundation under the Trilateral Partnerships grant No. 90411. The authors acknowledge financial support by the Russian Basic Research Foundation, grants
15-52-12387, 16-02-00649, and by Department of Physical Sciences of RAS, subprogram `Interstellar and intergalactic media: active and elongated objects'.

\appendix

\section{The `proof' of the instability using Lyapunov function for $\lowercase{l} \gg 1$}

In this Appendix we reproduce the original proof by Antonov, made with the aid  of Lyapunov function for the case of large $l$, but in the notations and terms adopted in the present work.

When the radial derivatives of the perturbed potential can be neglected compared to the angular derivatives,
$$
  \p \Phi_1/\p r\ll (1/r)\,\p\Phi/\p\theta
$$
or
$$
  \frac{d^2\chi}{d r^2} \ll \frac{l^2}{r^2}\,\chi\ ,
$$
eqs. (\ref{eq:AB2a}) and (\ref{eq:AB2b}) can be simplified as
\begin{align}
  \frac{\p {\cal A}}{\p t}+{\hat D}{\cal A}+\frac{l^2}{r^2}\,{\cal B}=0\quad \textrm{and} \label{ap-a:A} \\
  \frac{\p {\cal B}}{\p t}+{\hat D}{\cal B}=\pi\,\phi(r)\,F_0(E)\ , \label{ap-a:B}
\end{align}
where ${\hat D}=\nu(E)\,\p/\p w\ \ \textrm{and} \ \nu(E)=\Omega_1(E,L=0)$. The Poisson equation (\ref{eq:chiw}) then can be reduced to an algebraic equation:
$$
  \chi(r)=-\frac{4\pi\,G}{l^2}\,r^2\,\Pi(r)\ , \ \ \ \Pi(r)=\frac{1}{r^2}\int {\cal
  A}\,dv_r\ .
$$
Now we introduce new variables $A$ and $B$:
$$
  {\cal B}=\pi\,F_0\,B\ , \quad {\cal A}=-l^2\pi F_0\,A\ ,
$$
and the systems (\ref{ap-a:A}) and (\ref{ap-a:B}) can be rewritten as
\begin{equation}
  \frac{\p B}{\p t}+{\hat D}\,B=(2\pi)^2 G \int F_0(E)\,A\, dv_r\quad \textrm{and}
  \label{ap-pB:}
\end{equation}
\begin{equation}
  \frac{\p A}{\p t}+{\hat D}\,A=\frac{B}{r^2}\ .
  \label{ap-A:}
\end{equation}
Following \citet{A73}, we construct the Lyapunov function
\begin{equation}
  L\!=\!\int\!\!\int\! dr\,dv_r\, F_0(E)\, A B\!=\!\int\! dw \int \frac{F_0(E)\,dE}{\nu(E)}\,A
  B\ .
  \label{ap-L:}
\end{equation}
Differentiating over time, one obtains:
$$
   \frac{dL}{dt}=\int\!\!\!\int dr\,dv_r\,
   F_0(E)\, \left(\frac{\p A}{\p t}\, B+\frac{\p B}{\p
  t}\,A\right)\ .
$$
Now rewriting (\ref{ap-pB:})
\begin{multline}
  \frac{\p B}{\p t}+{\hat D}\,B=(2\pi)^2 G  \\
  \times\int dr'\,\delta(r'-r)\,\int F_0(E')\,A(r',v_r')\, dv_r'\ ,
  \label{ap-ppB:}
\end{multline}
we have
\begin{multline}
   \frac{dL}{dt}=\int\!\!\int dr\,dv_r\, F_0(E)\, \left\{\left(-{\hat
  D}\,A+\frac{B}{r^2}\right)\, B\right.\\
  +\left.\left[-{\hat D}\,B\!+\!(2\pi)^2 G\!
  \int\! dr'\,\delta(r'\!-\!r)\!\!\int\! F_0(E')\,A(r',v_r')\, dv_r'\right]A\right\}.
  \nonumber
\end{multline}
Since $B\,({\hat D}\,A)+A\,({\hat D}\,B)$ is a full derivative over $w$,
$$
  B\,({\hat D}\,A)+A\,({\hat D}\,B) ={\hat D} (AB)=\nu(E)\,\frac{\p (AB)}{\p w}\ ,
$$
the corresponding integral
$$
  \int dr\int dv_r\,(...) = \int\nu^{-1}(E)\,dE\,\int dw\,(...)
$$
vanishes and we obtain
\begin{multline}
   \frac{dL}{dt}=\int \int dr\,dv_r\, F_0(E)\, \frac{B^2}{r^2} \\ +(2\pi)^2 G
  \int\!\! \int\! dr\,dr' \delta(r'\!-\!r) \\ \times
   \int F_0(E')\,A(r',v_r')\, dv_r' \int F_0(E)\,A(r,v_r)\,dv_r\ ,
  \nonumber
\end{multline}
or finally
\begin{multline}
   \frac{dL}{dt}=\int \int dr\,dv_r\, F_0(E)\, \frac{B^2}{r^2} \\
   + (2\pi)^2 G  \int dr \left[\int F_0(E)\,A(r,v_r)\,dv_r\right]^2.
  \label{ap-a:A9}
\end{multline}
The last equation is the full analog of the Antonov's expression for $dF/dt$ (but expressed  in our variables).\footnote{
Note that in the cited paper by \cite{A73} this expression (following eq. (7))
contains a misprint: the second term in the r.h.s.
of the expression for $dF/dt$ should
read:
$ 2\pi G \int r^2\,dr \left[\int dE_0\rho_{E_0}\,(\xi_++\xi_-)\right]^2$.}

The proof is based on the evident positiveness of both terms in (\ref{ap-a:A9}). However, as we already noted in the main text, the first term diverges at $r=0$, so rigorously speaking such a proof of the radial orbit instability is invalid.

\section{Integral equations in the limit of small $L_T$ (even $\lowercase{l}$).
Clarifying a sense of diverging coefficients $\lowercase{p_k}$}

In this appendix we restrict ourselves to the relatively compact derivation for the case of dispersed Agekyan model ($q=-1$), although generalisation to arbitrary $F(E)$ is possible. Besides, we shall consider even spherical harmonics $l$ only, but the desired relations used for interpretation of diverging integrals in the delta function technique are universal and valid for odd $l$ as well. So we assume
\begin{equation}
  F_0(E)=\frac{K(L_T)}{8\pi^3}\,\delta(E)\ ,
\end{equation}
where
\begin{equation}
  \lim\limits_{L_T\to 0}K(L_T)=\Omega\equiv\Omega_1(0,0)\approx 2.16\ ,
\end{equation}
and starting from the integral equation in the Lagrange form,
\begin{multline}
  \phi_{\,l_1,\,l_2}(E,L) =-\frac{4\pi
  G}{2l+1}\sum\limits_{l_1'=-\infty}^{\infty}\sum\limits_{l_2'=-l}^l
  D_l^{l_2'} \int dE'\,\int dL'\,  \\
 \times F(E',L')  \left[\dfrac{\p }{\p E'}\,
  \Omega_{l_1'l_2'}(E',L')+l_2'\,\dfrac{\p
  }{\p L'}\,\right]\, \\
 \times \frac{L'}{\Omega_1(E',L')}\,\frac{\phi_{\,l_1'\,l_2'}(E',L')
  \,\Pi_{l_1,\,l_2;\,l_1',\,l_2'}(E,L;E',L')}
  {\omega-\Omega_{l_1'l_2'}(E',L')}\ .
  \label{ap-b:ie_lf}
\end{multline}
Here we denote
\begin{equation}
  \phi_{l_1\,l_2}(E,L)=\frac{1}{\pi}\int\limits_0^{\pi}\cos\Theta_{l_1
  l_2}(E,L;w)\, \chi\bigl[r(E,L,w)\bigr]\,dw\ ,
  \label{ap-b:phi}
\end{equation}
where $\Theta_{l_1l_2}(E,L,w)$ is an angle,
\begin{equation}
  \Theta_{l_1\,l_2}(E,L;w)=
  \bigl(l_1+l_2\,\frac{\Omega_2}{\Omega_1}\bigr)\,w
  -l_2\delta\varphi\,(E,L;w)
\end{equation}
and
\begin{multline}
  \delta\varphi(E,L,w) =\frac{L}{\Omega_1}\int\limits_0^w
  \frac{dw'}{r^2(w')}  \\ = L\int\limits_{r_{\rm min}(E,\,L)}^{r(E,L,w)}
  \frac{dx}{x^2\,\sqrt{\phantom{\big|}[2E+2\Psi(x)]-L^2/x^2}}
\end{multline}
is the azimuthal angle as the particle travels from pericentre to current radius $r$, $\Psi$ is the relative potential, $\Psi(r) \equiv - \Phi_0(r)$. In particular, in the apocentre ($w=\pi$) this angle is $\delta\varphi(E,L;\pi)=(\Omega_2/\Omega_1)\,\pi$. The kernel functions are
\begin{multline}
  \Pi_{l_1,\,l_2;\,l_1',\,l_2'}(E,L;E',L')=\oint dw
  \cos\Theta_{l_1\,l_2}(w)\\ \times
  \oint dw'\cos\Theta_{l_1'\,l_2'}(w') \,{\cal
  F}_l(r,r') \\
   =4\int\limits_{0}^{\pi} dw
  \cos\Theta_{l_1\,l_2}(w)\int\limits_0^{\pi}
  dw'\cos\Theta_{l_1'\,l_2'}(w')\,
   \,{\cal F}_l(r,r')\ .
  \label{ap-b:kf}
\end{multline}
Note that the symmetry of the radial function $r(2\pi- w) = r(w)$ allows one to reduce integration in eqs (\ref{ap-b:phi}) and (\ref{ap-b:kf}) over full range of the angle variable to the interval $[0,\pi]$.

The r.h.s. of (\ref{ap-b:ie_lf}) can be divided into two parts
\begin{equation}
  \phi_{l_1l_2}(0,L_T)=Q_E+Q_L\ ,
\end{equation}
where
\begin{multline}
  Q_E=-\frac{K(L_T)}{2\pi^2\,(2l+1)\,L_T^2}\sum\limits_{l_1'=-\infty}^{\infty}
  \sum\limits_{l_2'=-l}^l D_l^{l_2'}
    \\ \times
   \left[\frac{\p}{\p E'}\int\limits_0^{L_T}
  L'dL'\,\frac{\Omega_{l_1'l_2'}(E',L')}{\Omega_1(E',L')} \right. \,
  \\ \times \left.
  \frac{\phi_{l_1'l_2'}(E',L')\,
  \Pi_{l_1,l_2;l_1'l_2'}(0,L;E',L')}{\omega-\Omega_{l_1'l_2'}(E',L')}\,\right]_{E'=0}
  \label{ap-b:Q1}
\end{multline}
and
\begin{multline}
  Q_L=-\frac{{ K(L_T)}}{2\pi^2(2l+1)\,L_T}\sum\limits_{l_1'=-\infty}^{\infty}
  \sum\limits_{l_2'=-l}^l (l_2'D_l^{l_2'})  \\ \times
  \frac{\phi_{l_1'l_2'}(0,L_T)\,\Pi_{l_1,l_2;l_1'l_2'}(0,L_T;0,L_T)}
  {\Omega_1(0,L_T)\,
  [\omega-\Omega_{l_1'l_2'}(0,L_T)]}\ .
  \label{ap-b:Q2}
\end{multline}

Now we shall expand the integral equation entities on the small parameter $L_T$. The linear combination of frequencies can be rewritten through the precession rate $\Omega_{\rm pr}$,
\begin{multline}
  \Omega_{l_1l_2}=l_1\Omega_1+l_2\Omega_2
  = (l_1+{\case{1}{2}}\,l_2)\,\Omega_1\\
  +l_2\,(\Omega_2-{\case{1}{2}}\,\Omega_1)
  = (l_1+{\case{1}{2}}\,l_2)\,\Omega_1+l_2\,\Omega_{\rm pr}\ .
  \label{ap-b:O}
\end{multline}
For $\Theta_{l_1\,l_2}(E,L;w)$ one can write
\begin{equation}
  \Theta_{l_1\,l_2}(E,L;w)=[(l_1+{\case{1}{2}}\,l_2)\,w-
  {\case{1}{2}}\,l_2\pi]+l_2\,\beta\ ,
  \label{ap-b:th}
\end{equation}
where
\begin{equation}
  \beta=\frac{\Omega_{\rm
  pr}}{\Omega_1}\,(w-\pi)+\frac{L}{\Omega_1}\int\limits_w^{\pi}\frac{dw'}{r^2(w')}\ ,
  \label{ap-b:a1}
\end{equation}
or
\begin{equation}
  \beta=\frac{\Omega_{\rm pr}}{\Omega_1}\,(w-\pi)+
  L\int\limits_r^{r_{\rm max}}
  \frac{dr'}{r'^2\,\sqrt{\phantom{\big|}[2E+2\Psi(r')]-L^2/r'^2}}\ .
  \label{ap-b:a2}
\end{equation}
Here terms proportional to $\Omega_{\rm pr}$ and $L$ are considered to be small and vanishing as $L$ approaches zero. To be clear, we assume the lower limit in the integral in  (\ref{ap-b:a1})  $(L/\Omega_1)\int_w^{\pi}{dw'}/{r^2(w')}$ is not too close to zero, otherwise this integral becomes of the order unity, since for $w=0$ it equals to $\pi\,(\Omega_2/\Omega_1)\approx \frac{1}{2}\,\pi$. However, the range of $w$ where the integral becomes $\sim 1$ is very small for $L \to 0$, and we shall see below that this bring no difficulties in further integrations. In (\ref{ap-b:th}) we take into account that the angle $\delta\varphi$ changes from zero to $\approx \pi/2$ in the centre, and then remains almost constant in the remaining part of the orbit. Angle $\beta$ is the remaining part of angle $\delta\varphi$ gained from $r$ to $r_{\rm max}$. Thus $\beta$ is small as long as $r\gg r_{\rm min}$ in (\ref{ap-b:a2}), and contribution to $\delta\varphi$ gained near the centre is taken into account by the term $-\frac{1}{2}\,l_2\,\pi$ in the square brackets in (\ref{ap-b:O}).

For the even $l$, values of $l_2$ in the integral equation are even, so the sum $l_1+{\case{1}{2}}\,l_2$ is an integer. Introducing new indices
\begin{equation}
  n=l_1+{\case{1}{2}}\,l_2\ ,\quad n'=l_1'+{\case{1}{2}}\,l_2'
\end{equation}
one can switch in expressions for $Q_1$ and $Q_2$ from double summation over $l_1'$ and
$l_2'$ to summation over $n'$ and $l_2'$,
\begin{equation}
  \Theta_{l_1l_2}\to\Theta_{n\,l_2}=(n\,w-{\case{1}{2}}\,l_2\pi)+l_2\,\beta\ ,
\end{equation}
\begin{equation}
  \Omega_{l_1l_2}\to n\,\Omega_1+l_2\Omega_{\rm pr}\ .
\end{equation}

For $\phi_{l_1l_2}$ one obtains, providing $\Omega_{\rm pr}$ and $\alpha$ are small,
\begin{equation}
  \phi_{l_1',\,l_2'}(E',L')=\left[\Phi_{n'}(E')- l_2'\,(\delta\Phi)_{n'}(E')\right]\,(-1)^{l_2'/2}\ ,
\end{equation}
where
\begin{equation}
  \Phi_{n'}(E')=\frac{1}{\pi}\,\int\limits_0^{\pi}\cos
  (n'\,w')\,\chi(r')\,dw'
  \label{ap-b:b16}
\end{equation}
and
\begin{equation}
  (\delta\Phi)_{n'}(E')=\frac{1}{\pi}\,\int\limits_0^{\pi}\sin
  (n'\,w')\,\beta(E',w')\,\chi(r')\,dw'\ .
  \label{ap-b:dPh}
\end{equation}
Similarly, for the kernel functions
\begin{multline}
  \Pi_{l_1 l_2;\,l_1'l_2'}(E,0;E',L')=
 [{\cal K}_{n,n'}(E,E') \\
 - l_2'\,(\delta{\cal
  K})_{nn'}(E,E')]\,(-1)^{l_2/2+l_2'/2},
\end{multline}

\begin{multline}
  {\cal K}_{n,n'}(E,E')=4\int\limits_0^{\pi}
  dw\,\cos(nw)\\ \times\int\limits_0^{\pi}
  dw'\,\cos(n'w')\,{\cal F}_l(r,r')
\end{multline}
and
\begin{multline}
  (\delta{\cal K})_{nn'}(E,E')=4\int\limits_0^{\pi}
  dw\,\cos(nw)
 \\ \times
  \int\limits_0^{\pi} dw'\, \sin(n'w')\,\beta(w')\,{\cal
  F}_l(r,r')\ .
  \label{ap-b:dK}
\end{multline}
Since $\beta(w')$ in (\ref{ap-b:dPh}) and (\ref{ap-b:dK}) is multiplied by $\sin(n'w')$, which vanishes at $w=0$, the uncertainty in $\beta$ at $w\approx 0$ does not lead to any difficulties.

Now it is easy to relate eqs. (\ref{ap-b:ie_lf}) and (\ref{eq:ierad}). In the leading order over $L$, $\phi_{l_1l_2}$ coincides with $(-1)^{l_2/2} \Phi_n$ and the kernel functions $\Pi_{l_1l_2;l_1'l_2'}$
coincide with ${\cal K}^{\rm even}_{nn'}\cdot (-1)^{l_2/2+l_2'/2}$ of eq. (\ref{eq:ierad}). Using the identities
\begin{equation}
  \sum\limits_{l_2=-l}^l D_l^{l_2}=1\ ,\ \sum\limits_{l_2=-l}^l l_2 D_l^{l_2}=0\ \textrm{and}\ \sum\limits_{l_2=-l}^l l_2^2 D_l^{l_2}=\frac{l(l+1)}2
  \label{ap-b:id1}
\end{equation}
one can show that $Q_E$ turns into the last term containing the energy derivative. The remaining term, $Q_L$, vanishes in the leading order ${\cal O}(1/L_T)$,
\begin{multline}
  Q_L=(-1)^{l_2/2} \,\Biggl[- \frac{{\bar K(L_T)}}{2\pi^2(2l+1)\,L_T}\sum\limits_{n'=-\infty}^{\infty}
  \sum\limits_{l_2'=-l}^l
  (l_2'\,D_l^{l_2'})\\
  \times
  \frac{\Phi_{n'}(E') \, {\cal K}_{n,n'}(E,E') \, }{\Omega\,
  (\omega-n'\Omega)}\Biggr] = 0
\end{multline}
because of the second identity in (\ref{ap-b:id1}). To proceed further, we have to expand $Q_L$ to the next order ${\cal O}(L_T^0)$ and compare it with the first square bracket in (\ref{eq:ierad}).

Small additional terms $(\delta\Phi)_{n'}$ and $(\delta{\cal K})_{nn'}$ can be expanded over functions of the leading order. According to (\ref{ap-b:b16})
\begin{equation}
  \chi(r)=\sum_k \Phi_k(E)\,e^{ikw}\ ,
\end{equation}
and from (\ref{ap-b:dPh}) one has
\begin{equation}
  (\delta\Phi)_{n}(E)=  \sum\limits_k \beta_{nk}(E)\,\Phi_k(E)\ ,
\end{equation}
where
\begin{equation}
  \beta_{nk}(E,L)=
  \frac{1}{\pi}\int\limits_0^{\pi}\sin(n\,w)\cos(kw)\,\beta(E,L,w)\,dw\ .
  \label{ap-b:ank}
\end{equation}
Similarly, for $(\delta{\cal K})_{nn'}$ one obtains
\begin{equation}
  (\delta{\cal K})_{nn'}(E,E')=\sum\limits_k \,{\cal K}_{nk}(E,E')\,\beta_{n'k}(E',L')\ .
\end{equation}
Summarising, for $L=0$, $L' \ll 1$ one obtains:
\begin{multline}
  \phi_{l_1'l_2'}(E',L')\to \\ \left[\Phi_{n'}(E')-l_2'\sum_m\beta_{n'm}(E',L')\,\Phi_m(E')\right]\,(-1)^{l_2'/2},
\end{multline}

\begin{multline}
  \Pi_{l_1l_2;l_1'l_2'}(E,L=0;E',L')\to \left[{\cal
  K}_{nn'}(E,E')  \phantom{l_2'\sum\limits_m \,{\cal
  K}_{nm}(E,E')\,\beta_{n'm}(E',L')}  \right. \\ - \left. l_2'\sum\limits_m \,{\cal
  K}_{nm}(E,E')\,\beta_{n'm}(E',L')\right]\,(-1)^{l_2/2+l_2'/2}
\end{multline}

\begin{multline}
  \frac{1}{\omega-\Omega_{l_1'l_2'}}\to
  \frac{1}{\omega-n'\,\Omega-l_2'\,\Omega_{\rm pr}} \\
  \approx \frac{1}{\omega-n'\,\Omega}+\frac{l_2'\,\Omega_{\rm
  pr}}{(\omega-n'\,\Omega)^2}.
\end{multline}

Using these expressions in (\ref{ap-b:Q2}), we obtain in the order ${\cal O}(L_T^0)$:
\begin{multline}
  Q_L\approx-\frac{(-1)^{l_2/2}}
  {2\pi^2(2l+1)\,L_T}\sum\limits_{n'=-\infty}^{\infty}
  \sum\limits_{l_2'=-l}^l (l_2'^2\,D_l^{l_2'})\, \\
  \times\left\{-\frac{1}{\omega-n'\Omega}\left[ K_{nn'}\sum\limits_m
  \beta_{n'm}\,\Phi_m+\Phi_{n'}\sum\limits_m\beta_{n'm}\,{\cal
  K}_{nm}\right] \right. \\
  +\left. \frac{\Omega_{\rm pr}\,{\cal
  K}_{nn'}\Phi_{n'}}{(\omega-n'\Omega)^2}\right\}.
  \label{ap-bQ:}
\end{multline}
Summing up over $l_2'$ with the help of (\ref{ap-b:id1}) allows us to reduce (\ref{ap-bQ:}) to
\begin{multline}
  Q_L \equiv {\bar Q}_L\, (-1)^{l_2/2}\approx \frac{(-1)^{l_2/2}}{4\pi^2}\,\frac{l\,(l+1)}
  {2l+1}\,\frac{1}{L_T} \\
  \times\sum\limits_{n'=-\infty}^{\infty}
  \left[\sum\limits_m
  \frac{\beta_{n'm}\,({\cal K}_{nn'}\,\Phi_m+{\cal
  K}_{nm}\,\Phi_{n'})}{\omega-n'\Omega}-\underbrace{\frac{\Omega_{\rm pr}\,{\cal
  K}_{nn'}\Phi_{n'}}{(\omega-n'\Omega)^2}}\right].
  \label{ap-b:Q2red}
\end{multline}
The expression ${\bar Q}_L$ should be compared with the first term  in  r.h.s. of Eq.  (\ref{eq:even_fin}), which can be rewritten as
 \begin{multline}
  {Q}_L^{\rm pure\
  radial}=-\frac{1}{4\pi^2}\,\frac{l\,(l+1)}{2l+1}  \\ \times
  \sum\limits_{n'}\,\Bigg[
   \sum\limits_{m\ne n'}
   \frac{p_{n'-m}\,({\cal K}_{n\,n'}\,\Phi_m+{\cal K}_{n\,m}\Phi_{n'})}{\Omega\,(n'-m)}\,
  \frac{1}{\omega-n'\,\Omega} \\
    + \underbrace{\frac{{\cal K}_{n\,n'}\Phi_{n'}\,p_0}{(\omega-n'\,\Omega)^2}}\Bigg].
  \label{ap-b:Q2pr}
\end{multline}
In particular, comparison of the underbraced terms in (\ref{ap-b:Q2red}) and (\ref{ap-b:Q2pr}) gives that $p_0$ should be associated with
$\lim\limits _{L_T\to 0}\Omega_{\rm pr}(L_T)/L_T$, i.e.
\begin{equation}
  p_0\equiv \frac{1}{2\pi}\oint \frac{dw}{r^2(w)}\to   \lim\limits_{L_T\to 0}\frac{\Omega_{\rm pr}(L_T)}{L_T}\ .
  \label{ap-b:p0}
\end{equation}
Next, from (\ref{ap-b:a1}) and (\ref{ap-b:ank})
\begin{multline}
  \beta_{nm}=-\frac{1}{2\,\Omega}\,\Bigl\{\frac{1}{n+m}\Bigl[\frac{L}{\pi}\int\limits_0^{\pi}
  \frac{\cos(n+m)w}{r^2}\,dw-\frac{\Omega}{2}\Bigr] \\
  +\frac{1}{n-m}\Bigl[\frac{L}{\pi}\int\limits_0^{\pi}
  \frac{\cos(n-m)w}{r^2}\,dw-\frac{\Omega}{2}\Bigr]\Bigr\}
\end{multline}
for $m\ne \pm n$ and
\begin{equation}
  \beta_{n,\pm
  n}=-\frac{1}{4\,\Omega\,n}\,\Bigl[\frac{L_T}{\pi}\int\limits_0^{\pi}
  \frac{\cos(2nw)}{r^2}\,dw-\frac{\Omega}{2}\Bigr]\,
\end{equation}
for $m=\pm n$. Then, introducing
\begin{equation}
  P_k(E,L_T)=\frac{1}{L_T}\,
  \Bigl[\frac{L_T}{\pi}\int\limits_0^{\pi} \frac{\cos
  (kw)}{r^2}\,dw-\frac{\Omega(L_T)}{2}\Bigr]\ ,
\end{equation}
one can have
\begin{multline}
  (\beta_{nm})_{m\ne \pm n}=-\frac{L_T}{2\Omega}\,\Bigl(\frac{P_{n+m}}{n+m}+\frac{P_{n-m}}{n-m}\Bigr)\ , \\
  \beta_{n,\pm n}=-\frac{L_T}{4\,\Omega\,n}\,P_{2n}\ .
\end{multline}
Now it is not difficult to show that ${\bar Q}_{L}$ completely coincides with ${Q}_{L}^{\rm pure\ radial}$ if one associate $p_k$ as limiting values of $P_k(L_T)$:
\begin{equation}
  p_k=\frac1\pi \int_0^{\pi}\frac{\cos(kw)\,dw}{r^2(E,w)} \to \lim\limits_{L_T\to 0}P_k(E,L_T).
  \label{ap-b:pk}
\end{equation}
In particular, for $k=0$ one obtains (\ref{ap-b:p0})
\begin{multline}
  p_0\equiv \frac{1}{2\pi}\oint \frac{dw}{r^2(w)}\to
  \lim\limits_{L_T\to 0}
  \frac{1}{L_T}\Bigl[\frac{L_T}{\pi}\int\limits_0^{\pi}
  \frac{dw}{r^2}-\frac{\Omega_1(L_T)}{2}\Bigr] \\
  =\lim\limits_{L_T\to 0}
  \frac{\Omega_2(L_T)-\frac{1}{2}\,\Omega_1(L_T)}{L_T}\equiv\lim\limits_{L_T\to
  0} \frac{\Omega_{\rm pr}(L_T)}{L_T}.
\end{multline}

From definition of $p_k$,
\begin{equation}
  p_k=p_0- \frac{2}{\pi}\int\limits_0^{\pi} \frac {\sin^2 (\frac{1}{2}\,k  w)}{r^2(w)}\,dw\ ,
\end{equation}
with the integral converging in the usual sense, thus $p_{k\ne 0} - p_0 = {\cal O}(1)$.

\end{document}